\documentclass[reprint,superscriptaddress,aps,prd,longbibliography]{revtex4-2}
\usepackage{natbib}
\usepackage{kpfonts}
\usepackage{amssymb}
\usepackage{graphicx} % Required for inserting images
\usepackage{hyperref}
\usepackage{amsmath}
\hypersetup{
    colorlinks=true,
    linkcolor=blue,
    filecolor=magenta,
    urlcolor=blue,
    citecolor=red,
    pdfpagemode=FullScreen,
}

\begin{document}
\title{Hawking radiation inside a charged, cosmological black hole}
\author{Devayani Ravuri}
 \email{Devayani.Ravuri@colorado.edu}
 \affiliation{%
 JILA and Department of Physics, University of Colorado, Boulder, Colorado 80309, USA
}%
\author{Tyler McMaken}
 \email{tcmcmaken@umary.edu}
 \affiliation{%
 JILA and Department of Physics, University of Colorado, Boulder, Colorado 80309, USA
}%
 \affiliation{%
 Department of Mathematics and Physics, University of Mary, Bismarck, North Dakota 58504, USA
}%
\date{\today}
\begin{abstract}
    We study the effective temperature, as a rate of gravitational redshift, of the Hawking modes perceived by a radially free falling observer at an arbitrary location in a Reissner-Nordstr\"{o}m-(anti-)de Sitter [RN(A)dS] spacetime. In particular, the behavior of the modes at the inner horizon, and therein the validity of the strong cosmic censorship conjecture under the effective temperature formalism, has been analyzed across the physically permissible parameter space of the RNdS metric. The modes perceived by observers of both positive and negative specific energies have been taken into consideration. Finally, the behavior of the adiabatic control function has been examined over the position space of the observer to determine the regimes where the effective temperature function yields a Planckian spectrum of thermal radiation.
\end{abstract}

\maketitle

\hbadness=99999

\section{Introduction}\label{intro}
Black holes emerged as one of the most intriguing results of general relativity, describing a region where the curvature of space time is so intense that nothing, not even light, can escape. Although classically, black holes must not emit anything, not doing so violates the second law of thermodynamics \cite{Bardeen_1973,Bekenstein_1973}. The entropy of the black hole requires that it have a temperature, and the temperature must be associated with an emitted radiation. The coupling of the quantum and gravitational fields, at the horizons of a black hole, manifests as a thermal radiation with a temperature given as $T_H=\frac{\varkappa_+\hbar}{2\pi ck_B}$, where $\varkappa_+$ is the surface gravity of the horizon \cite{Hawking_1975}. Although Hawking's calculations of the temperature of the radiation were limited to observers asymptotically far from the black hole, attempts have since been made to extend the results for more general observers. 

The formulation relevant to this paper is based upon the result that an exponential relation between the affine parameters (\textit{u},\textit{U}) of past and future null infinities results in Hawking radiation \cite{Barcel__2011,Liberati_2011,Carballorubio_2024}. A function $\kappa(\textit{u})$ is defined, dependent on the parametrized relationship \textit{U}(\textit{u}), that primarily governs the peeling properties of the null geodesics \cite{Cropp_2013}. It has been shown that given an exponential relation between \textit{u} and \textit{U}, and given that the adiabatic condition [Eq.~\eqref{eq:adiabaticcondition}] holds, a thermal flux will be generated, whose temperature is proportional to $\kappa(\textit{u})$. In the limit of a static black hole, $\kappa$ becomes equal to the surface gravity $\varkappa_+$ of the event horizon. It is possible to recast the definition of the effective temperature $\kappa$ as the rate of exponential redshift of the Hawking modes. In the case of a collapsing mass, the modes experience a blueshift as they approach the mass and are subject to a gravitational redshift as they climb out. Another explanation that well complements this rather nonintuitive formalism notes that the vacuum state in which the thermal radiance is observed is related to the local Minkowski frame of the observer via an exponential scale transformation \cite{hu_1996}. This inherently draws upon the analogous relation between the Unruh effect and the Hawking radiation\textemdash an accelerated observer in the Minkowski  spacetime, and observer at a constant distance from the black hole will both detect a flux of particles \cite{Unruh_1976}.

Further validation of the above formalism came from its successful application to the Hawking temperatures for a Schwarzschild black hole of mass $m$, predicting a temperature of $(1-\frac{2Gm}{c^2 r})^{-\frac{1}{2}}\frac{\hbar c^3}{8\pi m G k_B}$ for a static observer at radius $r$, which at infinity reduces to $\frac{\hbar c^3}{8\pi m G k_B}$, consistent with Hawking's calculations \cite{Barbado_2011}. Subsequently, the formalism was extended to a Reissner-Nordstr\"{o}m black hole \cite{McMaken_2023}, the complexity of which lies in the presence of the inner (Cauchy) horizon\textemdash a null hypersurface of infinite blueshift, beyond which predictability breaks down. The Hawking temperatures were found to diverge to negative infinity at the horizon, indicative of an infinite rate of blueshift. Israel and Poisson were the first to carry out a full nonlinear analysis of the backreaction induced by the instability upon the internal geometry of the black hole \cite{Poisson_1990}. The unbounded inflation of the mass parameter, resulting from the backscattered radiation became known as ``mass inflation''. Of particular interest, in the context of mass inflation, is that the slightest external perturbation induces a divergence in the curvature as the Cauchy horizon is approached. The strong cosmic censorship (SCC) conjecture \cite{Penrose_1969} cites this fact to contend that the Cauchy horizon cannot form in the presence of external perturbations. However, it had been observed that the incorporation of a positive cosmological constant will cause a redshift, that could, to some extent, negate the effects of the infinite blueshift \cite{Mellor_1990,Brady_1998,Brady_1992}\textemdash thereby allowing the Cauchy horizon to exist without the evolution of a singular entity. Classically, it appeared that charged black holes in de Sitter space, with the appropriate range of mass and cosmological parameters violated the censorship. A recent paper \cite{Hollands_2020} resolved the apparent discrepancy to a great extent, by analyzing the quantum stress-energy tensor at the Cauchy horizon for a Reissner-Nordstr\"{o}m-de Sitter (RNdS) black hole, which effectively permits a sufficient divergence in the radiation for a singularity to evolve. We will calculate the effective temperatures as a rate of redshift for an RNdS black hole, for radial infallers, the results of which support the SCC conjecture for subextremal black holes, but suggest a violation at certain extremal bounds.

Of significant consequence in the study of an RNdS black  hole, is the existence of a cosmological horizon (at radius $r_c$), a boundary that causally separates an observer inside ($r<r_c$) from the external universe ($r>r_c$). It is characteristically very similar to the event horizon, following the same classical and thermodynamic laws, therein radiating particles with a thermal spectrum \cite{Bardeen_1973}. A rigorous study on the thermodynamic properties of the cosmological horizon was presented by Hawking and Gibbons \cite{Gibbons_1977}, extending to particle creation in de Sitter spaces.  While the matter of particle creation in de Sitter space had for long been a topic of study, it had always been viewed in an observer-independent frame. If particle creation were observer independent, and the invariance of the de Sitter group \cite{Aldrovandi_2007,Cacciatori_2008,Araujo_2024} were to hold, an observer Lorentz-boosted into a new frame will see exactly the same spectrum as they would have before the boost. This cannot be true unless the rate of particle production was either zero or infinity, zero being the favored choice of the two. But no particle production would imply no thermal radiation emanating from the cosmological horizon. However it was shown that an observer accelerating in Minkowski spacetime will detect particle production, indicating that particle production must be dependent on the worldline of the observer. An observer moving on a timelike geodesic will detect an isotropic thermal radiation of temperature $T_H=\frac{\varkappa_c\hbar}{2\pi ck_B}$, where $\varkappa_c$ is the surface gravity of the cosmological horizon. Once again, the invariance of the de Sitter group requires that any other observer also on a timelike geodesic detect the same radiation\textemdash indicating that particle production is in fact observer dependent. We will arrive at the same conclusion, albeit via a different approach. Both the outgoing and the ingoing modes tend to the same isotropic temperature asymptotically far from the black hole, independent of the observer's specific energy. We will further extend the analysis to negative observer energies and a negative cosmological constant. 

The layout of the paper is as follows: Section~\ref{sec:formalism} introduces the adiabatic approximation, which will be the foundation of the paper, followed by the RNdS geometry and the construction of the vacuum state. Section~\ref{sec:efftemperature} derives the effective temperature functions from the adiabatic approximation for a positive cosmological constant (RNdS) and analyzes them for observers of both positive and negative specific energies. Section~\ref{sec:AdSTemp} goes through a similar analysis for observers in a contracting universe, with a negative cosmological constant (RNAdS). Finally, in Section~\ref{sec:Adiabaticity} the validity of the adiabatic approximation is examined to understand the extent to which the formalism yields a thermal spectrum.

Although expressions in the introduction follow the SI unit system, a convention of natural units $G=\hbar=k_B=c=1$ and metric signature $(-,+,+,+)$ will be maintained hereon. 

\section{Formalism} \label{sec:formalism}
\subsection{Geodesics in the RNdS geometry}
The line element in Cartesian coordinates for a static, spherically symmetric charged black hole in asymptotically de Sitter space is
\begin{equation}\label{eq:LineElement}
    ds^2=-\Delta(r)dt^2+\frac{dr^2}{\Delta(r)}+r^2(d\theta^2+\sin^2(\theta)d\phi^2),
\end{equation}
where the horizon function
\begin{equation} \label{eq:horizonfn}
    \Delta(r)=1-\frac{2m}{r}+\frac{q^2}{r^2}-\frac{\Lambda}{3}r^2
\end{equation}
is a quartic, whose roots describe the positions of the horizons. Three positive roots correspond to the inner, event and cosmological horizons. The fourth, negative root, is not physically significant. The properties of the black hole are incorporated into the line element via the horizon function\textemdash with $m$ and $q$ being the mass and charge of the black hole respectively. $\Lambda$ is the cosmological constant describing expansion $(\Lambda>0)$ or the contraction $(\Lambda<0)$ of the universe. It is the effect of the cosmological constant that the spacetime is not asymptotically Minkowskian\textemdash which would require that the horizon function tend to one infinitely far from the black hole.

The temporal component of the four velocity of a radially free-falling observer is
\begin{equation}
    u_t=-E,\qquad u^t=\frac{dt}{d\tau}=\frac{E}{\Delta(r)}.
\end{equation}
Here, we have defined the specific energy of the observer as being the covariant temporal component of the observer's four velocity. This quantity is constant, and in a way, characterizes the geodesic. 

The components for the four velocity of an observer ($u_{\mu}$) and of the null particle ($k^{\mu}$) are found using the standard conditions

\begin{equation}
    -\Delta(r){(u^{t})}^2+\frac{1}{\Delta(r)}{(u^{r})}^2 = -1,
\end{equation}
\begin{equation}
    -\Delta(r){(k^t)}^{2}+\frac{1}{\Delta(r)}{(k^r)}^{2}=0.
\end{equation}
The frequency of the null particle as seen by the observer may be obtained by transforming the wave four vector of the null particle into the frame of the observer
\begin{equation}
    \omega=-k^\mu u_\mu.
\end{equation}
Normalized to the frequency $\omega_0$ perceived by an observer at rest, at a point $r_0$, 

\begin{equation}\label{eq:omega}
    \frac{\omega}{\omega_0}=\frac{E\mp \text{sgn}(u_r)\sqrt{E^2-\Delta}}{\Delta},
\end{equation}
where the upper (lower) signs correspond to outgoing (ingoing) modes respectively. $E$ is the specific energy of the observer, which remains constant along the natural geodesic. It was defined earlier as the negative temporal component of the velocity four vector. The value of the specific energy can be easily calculated as the square root of the horizon function at a point where $u_r=0$. We can conveniently choose (as we subsequently will) the rest point of the massive particle to be at the point where the horizon function is at a maximum. 

We will define the specific energy to be positive for an observer who starts out from rest at a point between the event and cosmological horizons. An infalling (outfalling) observer may be defined as one who travels to a smaller (larger) spatial radius by virtue of the natural geodesic trajectory. An ingoer (outgoer) is defined by their intended direction of motion. The distinction may not appear to be of great significance in a nontrapping region, where an observer can accelerate enough to be able to oppose the natural geodesic. In a trapping region, such as inside the event horizon, it is not possible to be able to travel to a larger spatial radius, and an attempted outward acceleration will induce the observer into a state of negative energy. An observer beyond the event horizon ($r<r_+$), who is infalling and outgoing, will have a specific energy $E<0$. So also, an observer beyond the cosmological horizon who is outfalling but ingoing will also have a negative specific energy. Note that the modes are defined as being either outgoing or ingoing, a state determined by the initial conditions at past null infinity. Outgoing modes inside the event horizon will remain outward directed, but will travel to a smaller spatial radius, eventually undergoing a divergence at the inner horizon. 

\subsection{Construction of the Unruh quantum state} \label{subsec: UnruhState}

In order to define the vacuum state of the quantum field in the spacetime geometry defined by Eq.~\eqref{eq:LineElement}, the massless scalar wave equation, $\mbox{$\square$}\phi_{\omega l m}=0$, may be solved. The field may be decomposed into a set of orthonormal positive and negative frequency modes

\begin{equation}\label{eq:modedecomposition}
    \phi=\int_{0}^{\infty} d\omega \sum_{l=0}^{l=\infty} \sum_{m=-l}^{m=+l} (a_{\omega lm}\phi_{\omega lm}+a^{\dagger}_{\omega lm}\phi^{*}_{\omega lm})\,.
\end{equation}

The modes can be further separated into radial and angular components,  of which the radial component, $f_{\omega l}$ must satisfy
\begin{equation}\label{eq:radialmodes}
    \frac{\partial^2 f_{\omega l}}{\partial r^{*2}} - \frac{\partial^2 f_{\omega l}}{\partial t^2} = \Delta \left(\frac{l(l+1)}{r^2}+\frac{1}{r}\frac{d\Delta}{dr}\right)f_{\omega l}.
\end{equation}
The tortoise coordinate $r^*$ is defined as
\begin{equation} \label{eq:tortoise}
    dr^*=\frac{dr}{\Delta}.
\end{equation}
The solutions to $f_{\omega l}$ yields two sets of waves\textemdash one coming in from infinity, the other emanating from the past horizon of the black hole, that is, the ingoing and outgoing modes respectively.

The quantum field's vacuum state is formally defined by canonical quantization of the ladder operators $a_{\omega lm}$ of Eq.~\eqref{eq:modedecomposition}, which practically manifests itself as the choice of past boundary conditions for the mode solutions $f_{\omega l}$ in Eq.~\eqref{eq:radialmodes}. The asymptotic past of a realistic black hole is flat everywhere, so that the vacuum state can be unambiguously defined by initializing a set of ingoing modes $f_{\omega l}=e^{-i\omega(r^*+t)}$ at infinity and propagating these modes through the collapsing matter to obtain outgoing modes. However, the RNdS metric used throughout this paper is eternal and static, and therefore it contains a horizon at $r_+$ in its asymptotic past boundary. The Unruh quantum state \cite{Unruh_1976} solves this problem by initializing a set of outgoing modes at the past horizon so that the resulting quantum modes throughout the spacetime are equivalent to those obtained by a collapse model taken far enough into the past.

A minor subtlety with the choice of the Unruh state to characterize the spacetime's semiclassical behavior is that the Unruh modes diverge everywhere along the Cauchy horizon, whereas in some shell collapse models \cite{Barcelo_2021}, the state can be made to remain regular as it passes through the outgoing portion of the inner horizon. However, regular evolution through the inner horizon is not expected to be generic at either the classical or semiclassical level as a result of the perturbation-induced mass inflation instability \cite{McMakenInstability_2023}. Thus, the Unruh state used here should be a reasonable approximation for the semiclassical late-time behavior of realistic black holes.

``Emitters'' will hereon represent the frame with respect to which the field is quantized. In an asymptotically Minkowskian spacetime, the ingoing modes are generated from a family of emitters in free fall at past null infinity. The modes are therefore positive frequency with respect to the proper time of a freely falling emitter in Minkowski spacetime. Similarly, the outgoing modes are generated from a family of freely falling emitters at the past horizon, and therefore defined to have a positive frequency with respect to the proper time of this family of emitters. ``Observers'', a term used previously, are simply representatives of an entity that has the ability to detect the radiation at arbitrary locations. 

The RNdS metric, however, is de Sitter, and not Minkowskian at past null infinity. Moreover, the cosmological horizon acts as a causal separation, so that a light ray originating from beyond will not be able to cross the horizon. The ingoing modes seen by an observer inside the cosmological horizon, must therefore be generated by emitters in free fall at the cosmological horizon (and not at infinity, as with other metrics). The boundary conditions must be defined at the past horizon and the cosmological horizon of the black hole. 

We will use a coordinate system defined in the RNdS spacetime as follows:
In terms of the tortoise coordinate, as defined in Eq.~\eqref{eq:tortoise}, the Eddington-Finklestein null coordinates are
\begin{equation}\label{eq:EFcoordinates}
    u=t-r^*,   
    \qquad
    v=t+r^*.
\end{equation}
In order to ensure continuity across the event and cosmological horizons, the Kruskal coordinates are defined as \cite{Hollands_2020}

\begin{subequations} \label{eq:KruskalC}
\begin{align}\label{eq:KruskalP}
     &U_+=-e^{-\varkappa_+u},\\
     &V_c=-e^{-\varkappa_c v}
\end{align}
\end{subequations}
respectively. $\varkappa_+=\frac{1}{2}\frac{d\Delta}{dr}\big|_{r_+}$
and $\varkappa_c=\frac{1}{2}\frac{d\Delta}{dr}\big|_{r_c}$
denote the surface gravities at the event at cosmological horizons respectively. 

In finding the normalized mode solutions for a spherically symmetric, static black hole, the angular components can be safely suppressed \cite{Barcel__2011}. This permits us to express the modes in a 1+1D spacetime. Additionally, the Polyakov approximation that permits a dimensional reduction of the quantum stress energy tensor for spherically symmetric, static and conformal metric, to a two-dimensional spacetime does not withhold any significant properties of the modes, in the geometric optics limit (which has already been assumed) \cite{McMakenInstability_2023}.
In this light, we may express the mode solutions in a 1+1D spacetime as
\begin{subequations}
\begin{align}
    &f^{in}_{\omega}=\frac{1}{\sqrt{4\pi\omega}}e^{-\iota\omega V_c},\\
    &f^{out}_{\omega}=\frac{1}{\sqrt{4\pi\omega}}e^{-\iota\omega U_+}.
\end{align}
\end{subequations}
Since modes generated from past null infinity undergo an exponential redshift as they traverse through the collapsing shell of matter, it is the high frequency modes that are of direct relevance to the calculation. Moreover, the high frequency modes are less subject to scattering. The choice of mode solutions as being proportional to the exponent of the proper time implicitly builds in the geometric optics approximation (also known as the high frequency approximation), thereby reducing a scattering problem to a ray tracing problem.
We will justify this construct by showing that the proper time of an emitter in free fall at the cosmological horizon is proportional to $V_c$, and likewise for an emitter at the event horizon. For convenience, and in accordance with the natural form of the metric, both the emitter and the observer will be designed to start from rest at a point in between the event and cosmological horizons. From that point, which is akin to an unstable equilibrium, they could either fall inward toward the event horizon or outward toward the cosmological horizon. It is important to note that a ``free falling'' emitter at the cosmological horizon, generating ingoing modes, is traveling to a larger spatial radius. The derivative of the outgoing null coordinate is:
\begin{equation}
    \frac{dv}{d\tau}=\frac{dt}{d\tau}+\frac{dr^*}{dr}\frac{dr}{d\tau}.
\end{equation}
In the limit as the emitter approaches the cosmological horizon,
\begin{equation}\label{eq:cosmologicallim}
    \lim_{r\to {r_c}} \frac{d\tau}{dv}\approx -\frac{(r_c-r)}{E}\varkappa_c,
\end{equation}
For an outfalling emitter, $r-r_c < 0$.
The radial component of the emitter's four velocity evaluated at the cosmological horizon gives
\begin{equation} \label{eq:ratio}
    \frac{dr}{d\tau}\bigg|_{r_c}=E.
\end{equation}
Then Eqs.~\eqref{eq:cosmologicallim} and \eqref{eq:ratio} together solve as
\begin{equation}
    \tau\propto e^{-\varkappa_c v}.
\end{equation}
But that is exactly the definition of the Kruskal coordinate $V_c$ along the cosmological horizon. A similar analysis can be done to show that the proper time of an emitter at the event horizon is proportional to $U_+$. 
The proper time of an emitter near the cosmological horizon decreases exponentially with the global time coordinate $t$, indicating an exponential redshift in the modes, and a positive effective temperature therein. It is important to note that this construction of the Unruh state is valid only for positive values of the cosmological constant. Further discussion on the vacuum states will be provided in the context of a contracting universe (see Section~\ref{subsec:AdSVacuum}).

\subsection{The effective temperature formalism}\label{subsec:adiabaticapprox}

Having defined both the geometry and the quantum state coupled to it, we may now offer a precise formulation of the effective temperature function.

The objective, as previously mentioned, is to generalize the calculation of the original Hawking temperature to arbitrary observers. The perception of thermal radiation, as shown in \cite{hu_1996}, arises from an exponential relation between the affine parameters of the emitter and the observer. Such an exponential relation is known to exist for a freely falling observer-emitter pair in a black hole metric (in which the observer detects a Hawking like radiation), or in a Rindler-Minkowski frame (in which the observer in the Rindler space sees an Unruh radiation).

If a Hawking flux is to be perceived, one can therefore define an exponential relation between the two affine parameters. Consider an emitter following a trajectory characterized by affine parameter $U$, and an observer following a timelike geodesic characterized by affine parameter $u$. The worldlines of the emitter and the observer are connected by a null ray following a geodesic path given by $U(u)$. Define a parametric relation $U(u)$ \cite{Barcel__2011}, via a function $\kappa(u)$:
\begin{equation} \label{eq:kappa}
     \kappa(u)=-\frac{d^2U/du^2}{dU/du}=-\frac{d}{du}\ln\left(\frac{dU}{du}\right).
\end{equation}
An integration of the above equation, to obtain $U(u)$ shows that this parametrization does indeed characterize an exponential relation. Physically, the adiabatic condition requires that around any given curve characterized by $u_*$, the value of the function $\kappa_*\equiv\kappa(u_*)$ remains approximately constant. Mathematically, the constraint can be given as
\begin{equation} \label{eq:adiabaticcondition}
    \left|\frac{1}{\kappa_*^2}\frac{d\kappa_*}{du}\right|\ll 1.
\end{equation}
Provided the adiabatic condition is satisfied, and in the geometric optics limit, the observer will perceive a Planckian spectrum of thermal Hawking-like radiation with an effective temperature given by 
\begin{equation}\label{eq:Hawking Temp}
    T_H=\frac{\kappa(u_*)}{2\pi}.
\end{equation}
Having scaled the constants, the effective temperature at any point can be well given by $\kappa(u_*)$. Of particular note is that this formalism is independent of the position $u_*$ of the observer. $\kappa(u)$ by definition, is the parametrized relation between the affine parameters of the emitter and the observer, describing the peeling properties of the null geodesics. It is on this quantity that the Hawking temperature is dependent, rather than the surface gravity itself. In the limit of a static black hole, i.e.,\ as the observer approaches future infinity, $\kappa(u)$ equals the surface gravity, reducing the expression to the well established Hawking temperature. This formulation Eq.~\eqref{eq:Hawking Temp} thereby permits a calculation of the effective temperature perceived by an observer at any point in spacetime. 

The affine parameter that characterizes the trajectory of any particle is the proper time. Therefore, the affine parameters \textit{u} and \textit{U} may be relabeled with the proper times $\tau_{ob}$ and ${\tau_{em}}$ of the observer and emitter respectively.
The effective temperature function Eq.~\eqref{eq:kappa} can be written as
\begin{equation} \label{eq:efftemp}
   \kappa(u)= -\frac{d}{d\tau_{ob}}\ln\left(\frac{d\tau_{em}}{d\tau_{ob}}\right) = -\frac{d}{d\tau_{ob}}\ln\left(\frac{\omega_{ob}}{\omega_{em}}\right).
\end{equation}
In this form, the effective temperature function is seen to be the rate of gravitational redshift. It is this form of the effective temperature function that will be relevant to the subsequent calculations.

\section{Effective temperature perceived by a radial free faller in RNdS spacetime(\texorpdfstring{$\Lambda>0$}{Lambda>0})}  \label{sec:efftemperature}
\subsection{The effective temperature functions} \label{subsec:efftempfunc}

The roots of the horizon function, given in Eq.~\eqref{eq:horizonfn} determine the positions of the horizons for a black hole of the determined parameters. The forms of these roots are neither analytically straightforward, nor particularly illuminating. A comprehensive description of the roots may be found in \cite{chrysostomou_2023}. To reduce the number of independent parameters in the horizon function, as in Eq.~\eqref{eq:horizonfn}, we define the de Sitter radius $L_{dS}$, related as $L_{dS}^2=\frac{3}{\Lambda}$. The mass and charge parameters ($m$ and $q$), are scaled by the de Sitter radius, and the new independent variables $M=\frac{m}{L_{dS}}$ and $Q=\frac{q}{L_{dS}}$ are employed in all the subsequent calculations.

Hereon, a more general form of the horizon function will be employed
\begin{equation}
    \Delta(r) = \frac{(r-r_-)(r-r_+)(r_c-r)(r-\Tilde{r})}{r^2 L_{dS}^2},
\end{equation}
where $r_-,\ r_+$ and $r_c$ are the inner (Cauchy), event, and cosmological horizons, respectively. The fourth root, $\Tilde{r}$ is negative and is not of great physical significance. The surface gravities take the form
\begin{subequations}
\begin{align}
    &\varkappa_- = \frac{(r_{-}-r_+)(r_c-r_-)(r_--\Tilde{r})}{2r_{-}^2 L_{dS}},\\
    &\varkappa_+ = \frac{(r_+-r_-)(r_c-r_+)(r_+-\Tilde{r})}{2r_{+}^2 L_{dS}},\\
    &\varkappa_c = -\frac{(r_c-r_{-})(r_c-r_+)(r_c-\Tilde{r})}{2r_{c}^2 L_{dS}}.
\end{align}
\end{subequations}
Surface gravities arise from the geodesic equations on a Killing horizon, and it is on this basis that they are defined. One may also define a ``generalized surface gravity'' \cite{Abreu_2010} anywhere in the spacetime for an observer on an imaginary surface with radius $r_{ob}$ as
\begin{equation}\label{eq:obsurfacegrav}
    \varkappa(r_{ob})=\frac{1}{2}\frac{d\Delta}{dr}\bigg|_{r_{ob}}.
\end{equation}

For radial trajectories, the effective potential takes the form $\frac{1}{2}(\Delta-1)$ \cite{Hartle_2003}. The horizon function therefore has a geometric construct similar to that of the potential function associated with the spacetime. Geometrically, the horizon function is always positive in between the event horizon and the cosmological horizon, which requires the existence of a maximum in the horizon function in the region\textemdash corresponding to an unstable equilibrium. The observer and emitter can be defined to be at rest at this point of maximum potential, so that their specific energy can be taken as the square root of the value of the horizon function at this point, and will remain constant as they traverse the geodesic. We will here on refer to this point as the rest point ($r_{0}$). An observer starting out from the rest point can either fall in toward the black hole, or outward toward the cosmological horizon and beyond. There will therefore be two categories of emitters/observers\textemdash the infallers and the outfallers. Ingoing modes are generated by a family of emitters (freely outfalling) at the past cosmological horizon, outgoing modes are similarly generated by a family of emitters in free fall (infalling) at the past event horizon. 

\begin{figure} [ht!]
    \centering
    \includegraphics[width=0.5\textwidth]{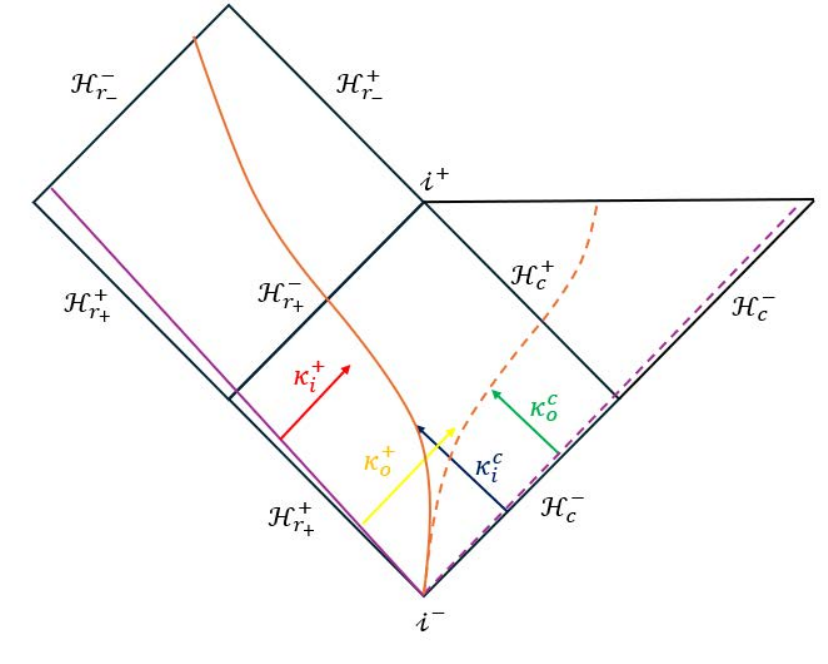}
    \caption{The orange (purple) lines represent the trajectories of the observer (emitter), and the solid (dashed) lines correspond to an ingoing (outgoing) trajectory. The red (yellow) and the blue (green) arrows represent the outgoing and ingoing modes respectively seen by an ingoing (outgoing) observer.}
    \label{fig:Penrose1}
\end{figure}

The effective temperature function [Eq.~\eqref{eq:efftemp}] can be expanded into the form
\begin{equation}
\begin{split} \label{eq:kappanew}
    \kappa&=-\frac{d}{d\tau_{ob}}\ln\left(\frac{\omega_{ob}}{\omega_{em}}\right)\\
    &=\mp\frac{1}{2}\frac{\omega_{ob}}{\omega_0}\left(\frac{d\Delta}{dr}\bigg|_{r_{ob}}-\frac{d\Delta}{dr}\bigg|_{r_{em}}\right),
\end{split}
\end{equation}
where the frequency perceived by the emitter or observer, normalized to the frequency perceived at rest $\omega_0$, at $r_0$ is given by Eq.~\eqref{eq:omega}. The upper (lower) signs refer to the outgoing (ingoing) modes.
We get a set of four effective temperature functions corresponding to ingoing/outgoing modes and ingoing/outgoing observers:
\begin{subequations} \label{eq:effectivetemp}
\begin{align}\label{eq:kappaip}
    &\kappa_i^{+}(r_{ob}) = -\frac{E+\sqrt{E^2-\Delta(r_{ob})}}{\Delta(r_{ob}) L_{dS}}\left(\varkappa(r_{ob}) -\varkappa_{+}\right),\\
    \label{eq:kappaic}
    &\kappa_i^{c}(r_{ob}) = \frac{E-\sqrt{E^2-\Delta(r_{ob})}}{\Delta(r_{ob})L_{dS}}\left(\varkappa(r_{ob}) -\varkappa_{c}\right),\\
    \label{eq:kappaop}
    &\kappa_o^{+}(r_{ob}) = -\frac{E-\sqrt{E^2-\Delta(r_{ob})}}{\Delta(r_{ob}) L_{dS}}\left(\varkappa(r_{ob})-\varkappa_{+}\right),\\
    \label{eq:kappaoc}
    &\kappa_o^{c}(r_{ob}) =\frac{E+\sqrt{E^2-\Delta(r_{ob})}}{\Delta(r_{ob}) L_{dS}}\left(\varkappa(r_{ob}) -\varkappa_{c}\right).
\end{align}
\end{subequations}
The superscript + (c) refers to the modes originating from the past event (cosmological) horizon, i.e.\ the outgoing (ingoing) modes. The subscript i (o) refers to an ingoing (outgoing) observer. 
Fig.~\ref{fig:Penrose1} shows the trajectories of the observer and the emitter, as well as the modes corresponding to the four effective temperatures in Eqs.~\eqref{eq:effectivetemp}. This being the Unruh state, the emitter is asymptotically close to the boundaries, and the observer can be either ingoing or outgoing from the point of maximum potential.

Though these effective temperatures may at first glance appear to diverge at the cosmological horizon due to the vanishing horizon function $\Delta(r_{ob})$ in the denominator of Eqs.~\eqref{eq:effectivetemp}, they nonetheless remain finite, since the numerator term $E-\sqrt{E^2 - \Delta(r_{ob})}$ also vanishes as $\mathcal{O}(\Delta)$ for outgoing modes, while for ingoing modes, the term $\varkappa(r_{ob})-\varkappa_c$ cancels out the blueshift divergence.
A preliminary analytical examination of the functions reveals that both $\kappa_o^+$ and $\kappa_o^{c}$ tend to constants at the cosmological horizon
\begin{equation} \label{eq: limitrc}
    \lim_{r\to r_c} (\kappa_o^+,\kappa_o^{c}) =\left( \frac{\varkappa_+ - \varkappa_c}{2E L_{dS}},\frac{E}{2\varkappa_c L_{dS}} \frac{d^2\Delta}{dr^2}\bigg|_{r_c}\right).
\end{equation}
A similar continuum in the modes can be seen across the event horizon, with the effective temperatures tending to values
\begin{equation}\label{eq:limitrp}
    \lim_{r\to r_+} (\kappa_i^+,\kappa_i^{c}) =\left(\frac{-E}{2\varkappa_+ L_{dS}} \frac{d^2\Delta}{dr^2}\bigg|_{r_+}, \frac{\varkappa_+ - \varkappa_c}{2E L_{dS}}\right).
\end{equation}
Were the observer to cross the cosmological horizon and proceed to infinity, the two modes will converge
\begin{equation}\label{eq:limitinf}
    \lim_{r\to\infty} (\kappa_o^{+}, \kappa_o^{c}) = \left(\frac{1}{L_{dS}},\frac{1}{L_{dS}}\right).
\end{equation}
In a purely de Sitter space, the ingoing and outgoing modes converge to $\frac{1}{L_{dS}}=\sqrt{\frac{\Lambda}{3}}$. In this form, the observed effective temperature in pure de Sitter space is recognizable as being equivalent to the magnitude of the surface gravity $\varkappa_c$ of the cosmological horizon \cite{Gibbons_1977}. 
At the inner horizon, the ingoing modes are finite, and equal to
\begin{equation}
   \lim_{r\to r_{-}} (\kappa_i^{c}) = 
   \frac{(\varkappa_{-}-\varkappa_{c})}{2E L_{dS}}.
\end{equation}
However, for outgoing modes, in the limit as $r\rightarrow r_-$, the effective temperature generally diverges, since the state-dependent surface gravity term $\varkappa(r_{ob})-\varkappa_+$ in Eq.~\eqref{eq:kappaip} generally does not vanish to cancel out the diverging blueshift term $(E-\sqrt{E^2-\Delta})/\Delta$. The effective temperature has first-order asymptotic form
\begin{equation} \label{eq:limrmkp}
    \lim_{r \to r_-}(\kappa^{+}_{i}) \approx \frac{E}{L_{dS}(r-r_-)}\left(\frac{\varkappa_+}{\varkappa_-}-1\right)-\frac{E}{L_{dS}}\frac{\varkappa'_-}{\varkappa_-}+\mathcal{O}(r-r_-),
\end{equation}
where $\varkappa'_- = \frac{d\varkappa}{dr}\big|_{r_-}$.
The exception (as will be subsequently seen) is if $\varkappa_- = \varkappa_+ (=0)$.
The above generalizations are subject to change in the context of an observer with negative specific energy (see Section~\ref{subsec:NegEnergy}).

\subsection{Effective temperature perceived by a radial freefaller with positive energy}

\begin{figure*}
    \centering
    
    \begin{minipage}{0.49\textwidth}
    \centering
        \includegraphics[width=1\textwidth]{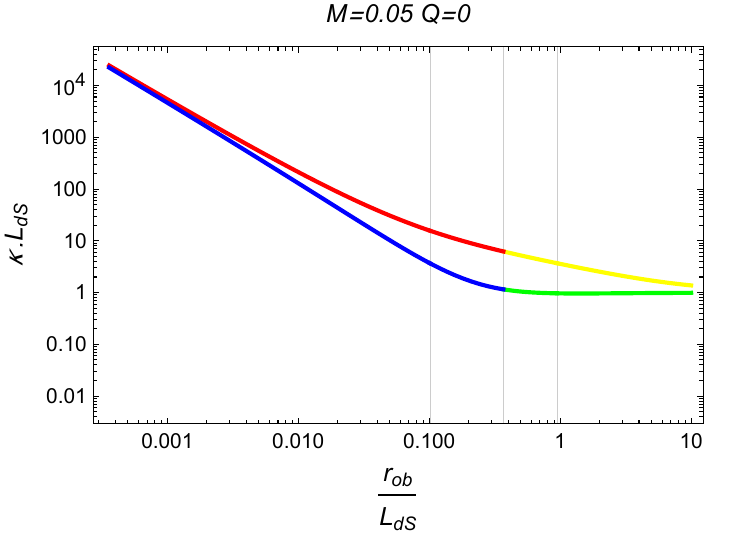}
    \end{minipage}
    \hfill
    \begin{minipage}{0.49\textwidth}
    \centering
        \includegraphics[width=1\textwidth]{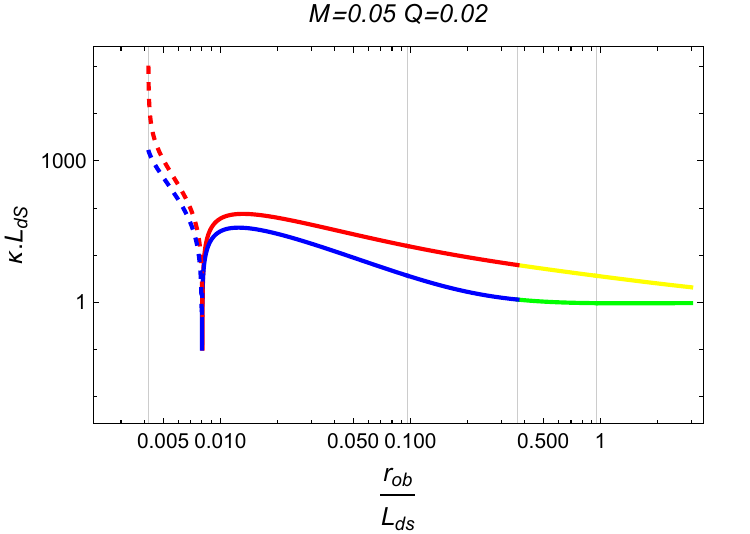}
    \end{minipage}%
    \vspace{1em}
    \begin{minipage}{0.49\textwidth}
    \centering
        \includegraphics[width=1\textwidth]{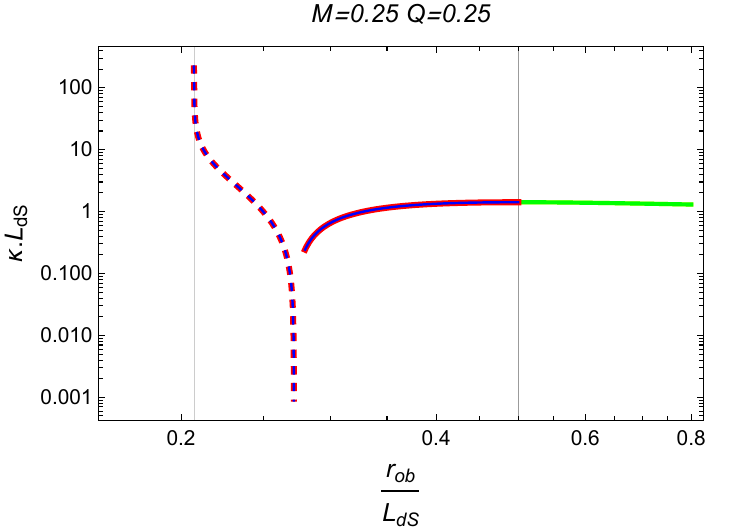}
    \end{minipage}
    \hfill
    \begin{minipage}{0.49\textwidth}
    \centering
        \includegraphics[width=1\textwidth]{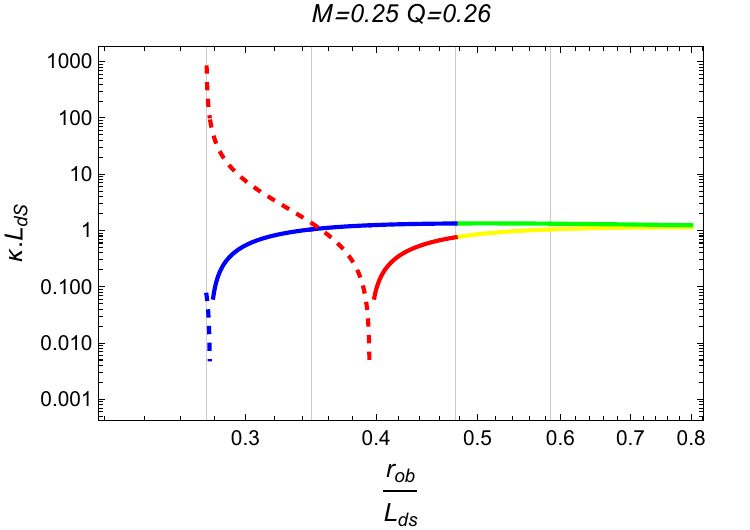}
    \end{minipage}
    
    \caption{Red (yellow) curves represent the effective temperature $\kappa_i^+ (\kappa_o^+)$ of outgoing modes for an ingoing (outgoing) observer, as a function of the position of the observer. Blue (green) curves are the effective temperatures $\kappa_i^{c} (\kappa_o^{c})$ for ingoing modes as seen by an ingoing (outgoing) observer. Grid lines are at the three horizons and at the rest point (between $r_+$ and $r_c$ where the blue and green lines intersect).}
    \label{fig:FIG 1}
\end{figure*}

Before proceeding to examine the extremal conditions of the RNdS metric, we may understand the features of the effective temperature in a Schwarzschild de Sitter (SdS) model, which is a simplification (\textit{Q}=0) of the same metric. It has only the event and cosmological horizons, thereby eliminating the intricacies that accompany the Cauchy horizon. It is therefore a good way to examine the individual effects of the mass and cosmological expansion on the effective temperature. The first graph, on the upper left of Fig.~\ref{fig:FIG 1}, shows the effective temperature for an SdS black hole as a function of the observer's position. Both the ingoing and outgoing modes are continuous across the event and cosmological horizons, in accordance with the Eqs.~\eqref{eq: limitrc} and \eqref{eq:limitrp}. This means that an ingoing(outgoing) observer will be able to cross the event(cosmological) horizon without noticing a discontinuity in the effective temperature. Neither the event nor the cosmological horizons act as a physical barrier, and are simply the radii at which space flows inward (or outward) at the speed of light. As the observer moves to a predominantly de Sitter (dS) space, the effective temperatures of the ingoing and outgoing modes converge to $1/L_{dS}$. In a purely dS space, then, the observer sees an isotropic equilibrium in both the modes, with the value of the effective temperature in this regime being independent of both the (positive) energy and the position of the observer. The energy-independence is in support of the observer-dependence of particle production in dS space \cite{Gibbons_1977}. The position-independence reinforces a homogeneity in the dS spacetime. At the other extreme, as the observer approaches the singularity, they will see a monotonic increase, and an asymptotic convergence in the positive effective temperatures of the modes. In the effective temperature formalism, a positive effective temperature corresponds to a perceived positive rate of redshift in the modes. As the curvature of spacetime increases toward the singularity, both the modes are pulled inward, and an infalling observer perceives an increasing rate of redshift. So also, the redshift in the modes in a purely dS spacetime may be understood as resulting from the increasing rate of the  expansion of space in this regime. The mass of the black hole and the cosmological expansion, individually, only cause a redshift. The blueshifting, that is seen in the other graphs of Fig.~\ref{fig:FIG 1}, must then be attributed entirely to the effects of the inner horizon.

A very typical representation of an RNdS black hole is seen in the upper right graph of Fig.~\ref{fig:FIG 1}, with parametric values of $M=0.05$, $Q=0.02$. Away from the inner horizon, it is fairly similar to the SdS case. Once again, there is a continuity across the event and cosmological horiozns, and the modes converge to unity in asymptotically dS space. The key feature of this graph, however, comes from the negative effective temperatures of the modes perceived by an observer approaching the inner horizon. A similarity in the cosmological and inner horizons stems from the fact that both have negative surface gravities. But unlike at the cosmological horizon, an observer who started from rest at the maximum of the horizon function and is subject to no external force agents, near the inner horizon is infalling (and ingoing). Effectively, the observer is falling inward and the modes are being repelled by the surface, resulting in a blueshift (which translates to a negative effective temperature). The transition from positive to negative effective temperatures must occur when the redshift caused by the event and cosmological horizons is overcome by the blueshift due to the inner horizon. This in turn, should have a dependency on the closeness of $r_+$ and $r_-$. Mathematically, the transition occurs at $\varkappa(r_{ob})=\varkappa_{+}$ for the outgoing modes and $\varkappa(r_{ob})=\varkappa_{c}$ for the ingoing modes. While the negative surface gravity certainly causes a blueshift, the effects are much stronger on the outgoing modes than on the ingoing modes because the repulsion supplements their intended direction. As a result, the outgoing modes diverge negatively, while the ingoing modes are blueshifted, but remain finite across the inner horizon. 

For a given value of the cosmological constant, the position of the transition point is entirely dependent upon the \textit{Q}/\textit{M} ratio. As the ratio increases, the effects of the inner horizon become predominant, and at a certain point, it becomes strong enough so that an observer outside the event horizon perceives a negative effective temperature in the outgoing modes. For small values of the cosmological constant, this critical ratio approaches $\sqrt{\frac{8}{9}}$, as is expected in the Reissner-Nordstr\"{o}m limit \cite{McMaken_2023}.  As the value of the cosmological constant increases, this critical ratio also increases; the effects of the inner horizon have to strengthen proportionally to combat the increasing redshift caused by the increasing cosmological constant. 
 The transition occurs exactly at the observer's rest point when $\frac{Q}{M}=\sqrt{\frac{9}{8}}$, which is the greatest ratio that would prevent a naked singularity \cite{PDavies_1989,chrysostomou_2023}, and corresponds to the degeneracy between the event and inner horizons. An ingoing observer would then only ever see modes that are blushifted. While this may not be of the greatest significance, it is interesting that the extremal criticality for the \textit{Q}/\textit{M} ratio in an RNdS black hole is exactly the reciprocal of the RN type black hole, where the effective temperatures become negative outside the event horizon for $\frac{Q}{M}>\sqrt{\frac{8}{9}}$. For the ingoing modes, the transition point gets closer to the inner horizon as the charge parameter reduces. In the SdS limit, there is no transition at all.
 
 An outgoing observer always perceives a convergence to $\varkappa_c$ in the effective temperatures of the ingoing and outgoing modes, in the infinite limit. The distance from the rest point at which the yellow and green curves converge is directly proportional to the \textit{M}/\textit{Q} ratio. For \textit{M}=\textit{Q}, the convergence is sharp, but only occurs asymptotically in the SdS limit. This is complemented by an analysis done in \cite{He_2018}, showing via an entropy analysis, that an SdS black hole is always thermodynamically unstable for all parametric values. 

\begin{figure} [ht!]
    \centering
    \includegraphics[width=0.5\textwidth]{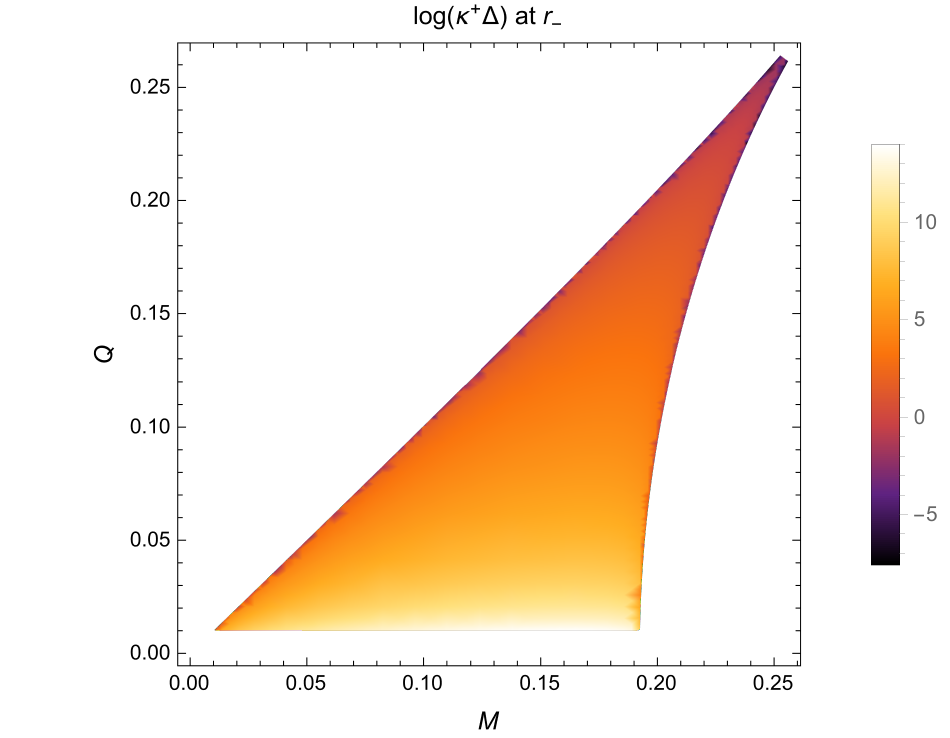}
    \caption{Density plot of $\log(\kappa^{+}_i\Delta)$ as a function of mass and charge in the limit as $r\xrightarrow{}r_-$. The plotted region demarcates the realistically permitted values for \textit{M} and \textit{Q}. A finite value of $\kappa^{+}\Delta$ as $\Delta \xrightarrow{} 0$ implies that $\kappa^+$ diverges. A highly negative value of $\kappa^{+}\Delta$ (the log tends to $-\infty$) indicates a finite effective temperature. }
    \label{fig:DensityPlotSCC}
\end{figure}

Having studied the general properties of a very typical RNdS black hole, we may now turn to the more extremal situations. A sharkfin diagram representing the physically permitted parameter space for the metric is plotted in Fig.~\ref{fig:DensityPlotSCC}. There are primarily, three regions of interest: the SdS limit (\textit{Q}=0), the charged Nariai limit ($r_+=r_c$) and the super-extremal limit ($r_+=r_-$).  In asymptotically flat spacetime, setting \textit{M}=\textit{Q} results in an extremal black hole, with a degeneracy in the event and inner horizons. The existence of the cosmological constant raises the upper bounds of the extremality, so that black holes may exist in a limited range where $Q>M$ (the cold region). \textit{M}=\textit{Q} (referred to as the lukewarm black hole family) is said to be one of the two conditions at which the black hole resides in a state of thermodynamic equilibrium \cite{chrysostomou_2023,Zhang_2016}. In the lukewarm state, the magnitudes of the surface gravities of the event and cosmological horizons are equal. However, equal surface gravities corresponds to equal Hawking temperatures. In the effective temperature formalism, an observer in a lukewarm metric will not notice an equilibrium in the effective temperatures of either the ingoing or the outgoing modes. The cosmological, event and inner horizons are still at distinct spatial radii. If, given that  $\varkappa_c=-\varkappa_+$, the observer were to perceive an equilibrium, one of the requirements would be that
\begin{equation}\label{eq:eqcondition}
    \frac{\Delta(r_{ob})}{\Delta_{0}}=\left(\frac{\varkappa_+-\varkappa(r_{ob})}{\varkappa_+}\right)\left(\frac{\varkappa_c-\varkappa(r_{ob})}{\varkappa_c}\right),
\end{equation}
with
$\varkappa(r_{ob})$ as defined in Eq.~\eqref{eq:obsurfacegrav}, and
$$\Delta_{0}=E^2.$$
This condition is not generally satisfied, except under certain regimes. An exception will be seen subsequently. 

The second region of thermal equilibrium, as stated in literature \cite{Zhang_2016,Romans_1992,Kastor_1996}, is the charged Nariai family. It is the extremality determined by the equivalence of $r_+$ and $r_c$. Furthermore, the surface gravities $\varkappa_{+}=\varkappa_{c}=0$, and the rest point of the observer coincides with these degenerate horizons. With the ingoing and outgoing modes being generated from exactly the same radius, and with the observer having started out from rest at exactly that same point, it is consistent that the rate of redshift/blueshift perceived by the observer is the same for both modes, resulting in the equilibrium.  This family demarcates the upper limit on the mass of the black hole, which naturally requires that the event horizon fit within the cosmological boundary of the universe. In an SdS universe, the Nariai branch has no singularities, but with the inclusion of the charge, an effective singularity forms at the inner horizon. The graph on the lower left of Fig.~\ref{fig:FIG 1}, with parameters $M=0.25$ and $Q=0.25$, falls at the intersection of the lukewarm and Nariai branches. It shows a state of complete thermodynamic equilibrium, where an observer anywhere in spacetime will see exactly the same effective temperatures if they looked at the sky above, or at the family of emitters below on the event horizon. Note that the equilibrium arises from the Nariai state, rather than from the lukewarm condition. With the event and cosmological horizons coinciding, the rest point of the observer is degenerate with the two horizons, so that the specific energy of the observer is also zero. In such a case, it is not hard to see that this is one of the few cases under which Eq.~\eqref{eq:eqcondition} must always hold true.
Following the limits of Eq.~\eqref{eq: limitrc}, the effective temperature in asymptotically dS spacetime tends to 1 (dS length units).
It has been calculated, in several papers \cite{Myung_2008,He_2018,chrysostomou_2023,Romans_1992}, that the effective temperature of the event and cosmological horizons goes to zero in the Nariai limit. It should be noted here, that the derivation of the ``effective temperature'' as in the above mentioned papers, stems from a thermodynamic perspective, unlike the present formalism. Regardless, their statement holds true given a standard normalization of the effective temperature. Our formalism inherently developed upon the Bousso-Hawking normalization \cite{Bousso_1996}, where the normalization constant ($\gamma_{t}$) in the timelike Killing vector ($\xi = \gamma_{t}\frac{\partial}{\partial t}$) is chosen to satisfy $\xi^{\mu}\xi_{\mu} = -1$ at the equilibrium point of the horizon function (which we defined to be the rest point of our observer) \cite{Eune_2013}.

The range in between the lukewarm and the $r_+ = r_-$ families, the extent of which is dependent on the strength of the cosmological constant, is the ``cold'' region. It is characterized by a charge marginally greater than the mass, as in lower right graph of Figure~\ref{fig:FIG 1}, given by parameters \textit{M}=0.25 and \textit{Q}=0.26. Once again, the transition for the outgoing modes occurs outside the event horizon (which is not unexpected, given that the charge is greater than the mass). The transition for the ingoing modes occurs very close to the inner horizon. The large values of both the mass and the cosmological will contribute a very strong redshift effect, and with the effects of the repulsion from the inner horizon being weaker on the ingoing modes, they remain positive in effective temperature until they get very close to the inner horizon. At the superextremal limit, where the inner and event horizons coincide exactly, the observer will never see ingoing modes of a negative effective temperature.

Finally, we may observe the effect of the magnitude of the cosmological constant upon the effective temperature. For a given \textit{M}/\textit{Q} ratio, the maximum positive value of the effective temperatures decreases as the cosmological constant is decreased. As the cosmological horizon gets pushed farther out, the maximum effective temperatures decrease. While effective temperature is certainly a different quantity from the rate of particle production, it is interesting to note that a study of the Schwinger effect on RNdS black holes suggested that the dS boundary pulls the event horizon toward the cosmological horizon and weakens the field on the event horizon, thereby reducing the particle pair production \cite{Chen_2020}. The de Sitter background suppresses particle production, while the AdS background enhances it. A similar conclusion was reached by Hawking and Bousso, by calculating the particle pair creation rate from instanton actions as a function of the cosmological constant \cite{Bousso_1996,Mann_1995}.

Having often used the term ``negative effective temperatures,'' it is worth taking a moment to consider the implications of a negative effective temperature and therein the importance of the transition points. Starting at the ground state, an increase in energy corresponds to the population of higher energy states, resulting in an increase in entropy, and this process can go on until the system reaches a state of maximum entropy (or infinite temperature). But the energy can be further increased if high energy states are more populated than the low energy states, a phenomenon commonly known as population inversion. At this point, the entropy decreases with an increase in energy, and results in a negative temperature. At maximum entropy, a discontinuity occurs as the temperature jumps from positive to negative infinity \cite{Braun_2013,Ramsay_1956,Norte_2024}. A paper in 1988 \cite{Pavon_1988}, perhaps one of the earliest on the topic of negative temperatures in a gravitational context, noted that self gravitating spheres of black body radiation could exhibit alternating regions of positive and negative heat capacities, with the discontinuity at infinite temperature being termed a ``phase transition''. Physically, negative temperatures are hotter than positive temperatures. In the context of a black hole, a negative effective temperature means that the observer will no longer perceive Hawking radiation as the Planckian spectrum of a blackbody. The exponential decay of the radiation at high frequencies instead becomes a divergence \cite{McMaken_2023pancakification}. 
 In our analysis, a negative effective temperature results from the blueshifting of the Hawking modes, that results from the presence of a horizon of negative surface gravity.

\subsection{Divergence at the inner horizon and on the validity of the SCC conjecture} \label{subsec:SCCviolation}

The strong cosmic censorship conjecture, in its precise definition \cite{PenroseGR_1980,Christodoulou_2008}, states that for generic asymptotically flat initial conditions for Einstein's equations, the maximal Cauchy development is inextendible. In cases of asymptotically flat spacetimes with a Cauchy horizon, this was easily understood in terms of the infinite blueshift in the radiation as it approached the inner horizon. This model had to be reanalyzed when a positive cosmological expansion was taken into account. The redshift contributed by an expanding universe could potentially negate the effects of the infinite blueshift at the inner horizon. The first comprehensive analysis of this effect was provided by Mellor and Moss \cite{Mellor_1990}, and subsequently expanded upon under different regimes of the de Sitter space \cite{Mo_2018,Cardoso_2018,Cardoso1_2018,Casals_2022,Destounis_2019,Guo_2019,Hintz_2017,Dias_2018,Dias1_2018}.

The solution to the charged Klein-Gordon equation gives the ingoing and outgoing modes, which can be approximated to the form
\begin{equation} \label{eq:modesols}
    \psi^{in} = e^{-\iota\omega u},
    \quad
    \psi^{out} = e^{-\iota\omega u}(r-r_-)^{\frac{i(\omega-\Phi(r_-))}{\varkappa_-}}
\end{equation}
near the inner horizon. The discontinuity in the outgoing modes was the basis for the SCC conjecture. The potential violation arises from parameters which permit a continuity in the modes, for which the mathematical requirement is that \cite{Dias_2018}
\begin{equation}
    \beta \equiv -\frac{\text{Im}(\omega)}{\varkappa_-} > \frac{1}{2}.
\end{equation}

Recent studies \cite{Hollands_2020,Hollands1_2020} have shown that for a 4 dimensional RNdS metric, even if there were parameters which classically permitted a continuity in the modes, the quantum stress-energy tensor $T_{\mu \nu}$ diverges provided that the initial state is regular near the initial Cauchy hypersurface. However, this divergence does not manifest as strongly in the 2 dimensional RNdS metric for $\varkappa_- = \varkappa_c$. But once again, the authors \cite{Hollands_2020,Hollands1_2020} argue that there are other means by which a divergence may be obtained in these exceptional cases. 
It appears to be sufficiently established that a violation of the SCC conjecture is not possible for any physical parameters of black holes.

The graphs from Fig.~\ref{fig:FIG 1} suggest a divergence in the outgoing modes at the inner horizon. But these graphs represent only four of an infinite number of possible combinations of \textit{M} and \textit{Q} over the entire physically permissible parameter space of the RNdS metric. Furthermore, although the outgoing modes certainly appear to diverge, there is no way, on the basis of the graphs alone, of conclusively determining whether they are diverging to infinity, or tending to a very large finite value. The sharkfin in Fig.~\ref{fig:DensityPlotSCC} tests this divergence at the inner horizon for the entire physically permissible parameter space, by plotting a density function of $\log(\kappa^{+}_i \Delta)$ over this region.

To understand the validity of the SCC conjecture under the effective temperature formalism (and therein the sharkfin), we will study the asymptotic forms of $\kappa^{+}_i$ [Eq.~\eqref{eq:limrmkp}] and the horizon function in the limit as the observer approaches the inner horizon. The horizon function takes the form
\begin{equation} \label{eq:limrmDelta}
   \lim_{r \to r_-} \Delta (r) \approx 2\varkappa_{-}(r-r_-).
\end{equation}
For typical black holes, Eq.~\eqref{eq:limrmkp} is dominated by the first term, which scales inversely with $(r-r_-)$, leading to a negative divergence. The product of $\kappa^{+}_i$ and $\Delta$, which are approximate inverses of each other, is therefore finite:
\begin{equation} \label{eq:limproduct}
    \lim_{r \to r_-} (\kappa^{+}_i \Delta) \approx \frac{2E}{L_{dS}}(\varkappa_+ - \varkappa_-).
\end{equation}
The converse must also be true (given that the asymptotic form of the horizon function will undergo no significant higher order corrections under extremal conditions)\textemdash if the product $\kappa^{+}_i \Delta$ is finite, the effective temperature of the outgoing modes must diverge as $(r-r_-)^{-1}$. This region of infinite divergence is shown in Figure~\ref{fig:DensityPlotSCC}, which can be seen to span the majority of the physically permissible parameter space, and all of the nonextremal regimes.

As the surface gravity of the inner horizon approaches zero (which occurs at the extremal cold limit), the second correction term in Eq.~\eqref{eq:limrmkp} starts to become predominant, and the surface gravity of the inner horizon may be corrected as $\varkappa_- \approx \varkappa'_{-}(r-r_-)$. The second term scales as $(r-r_-)^{-1}$, and tends to $+\infty$\textemdash countering the divergence of the first term to $-\infty$. The effective temperature, in this limit is finite as can also be seen from Eq.~\eqref{eq:kappaip}. The product $\kappa^{+}_i \Delta$ tends to zero. This can be seen as the dark line on the extremal cold limit of the sharkfin (close the $M=Q$ region). It would therefore appear that the SCC conjecture is being violated in the extremal cold regime. 

The Nariai limit ($r_+=r_c$) represented by the right curved edge on the sharkfin, maintains the divergence in the effective temperature. As the charge increases, the magnitudes of both $\varkappa_+$ and $\varkappa_-$ decreases (and therein the difference between them), the magnitude of the product reduces, and the curve starts to darken. At the ultracold limit ($r_-=r_+=r_c$ at $M=\sqrt{2}/3\sqrt{3}$), the product goes to zero, and the logarithm of the product tends to $-\infty$. In the SdS limit, the surface gravity of the inner horizon approaches $-\infty$, so that the right side of Eq.~\eqref{eq:limproduct} approaches $+\infty$. This can be seen in the sharkfin as the increasingly whiter region toward the lower edge.

In both the above cases, the limit of extremality has been discussed. Extremal black holes are topologically distinct from nonextremal black holes. One may therefore question the validity in considering the quantum field of an extremal black hole as being a continuity of the nonextremal case. Some papers \cite{Liberati_2000} have clearly claimed that the extremal case in no sense represents a limit of the nonextremal case but implies a real discontinuity. Under certain conditions, a continuity may be maintained as shown in \cite{Balbinot_2007}, but this may not be generically true. The discussion above should be understood as an (analytical) analysis of subextremal black holes as they approach asymptotically close to the extremal limit. The existence of purely extremal black holes is curtailed by the third law of thermodynamics, and they do not have a realistic channel of formation.

Attempts have been made to calculate $\beta$ numerically for quasinormal modes in the permitted RNdS spacetime \cite{Dias1_2018,Cardoso_2018}, and regions have been found where it does exceed $1/2$. On this basis, the possibility of a nondivergent behavior was postulated for three conditions: if $M=Q$ \cite{Dias_2018,Cardoso_2018}, in the limit of $\varkappa_- = \varkappa_c$ \cite{Hollands_2020}, or at the extremal cold limit. The effective temperature formalism upholds the SCC conjecture in the lukewarm limit, but fails in the other two. It is able to limit the regions of possible violation to the condition $\varkappa_+=\varkappa_-=0$, which is a step further than what the calculation of $\beta$ had predicted. Evidently, formulating the effective temperature as a rate of gravitational redshift is able to account for factors that $\beta$ cannot. But a conclusive identification of these factors will perhaps require a more detailed investigation into  the relation between $\beta$ and $\kappa$.

\subsection{Effective temperature perceived by a radial freefaller with negative specific energy} \label{subsec:NegEnergy}

An observer inside a trapping horizon, who continuously accelerates outward, against the curvature of spacetime acquires a negative energy (state with $\frac{dt}{d\tau}<0$). It is not possible to be outfalling inside a black hole, and therefore, the observer will still be traveling to a smaller spatial radius. But the outward acceleration will cause the observer to pass through the future Cauchy horizon, in contrast to an observer of positive energy, who will travel to the past Cauchy horizon. The outward acceleration by the observer, or the application of an external force may be seen as agents to Lorentz-boost the observer into a frame of negative energy.

\begin{figure} [ht!]
    \centering
    \begin{minipage}{0.49\textwidth}
    \centering
        \includegraphics[width=1\textwidth]{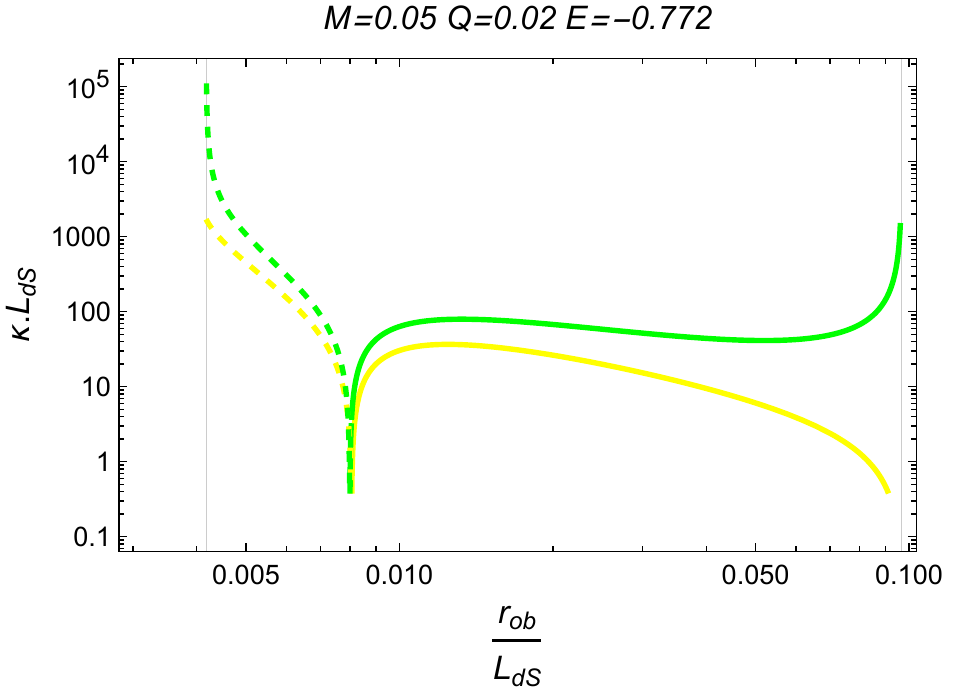}
    \end{minipage}%
    \vspace{1em}
    \hfill
    \begin{minipage}{0.49\textwidth}
    \centering
        \includegraphics[width=1\textwidth]{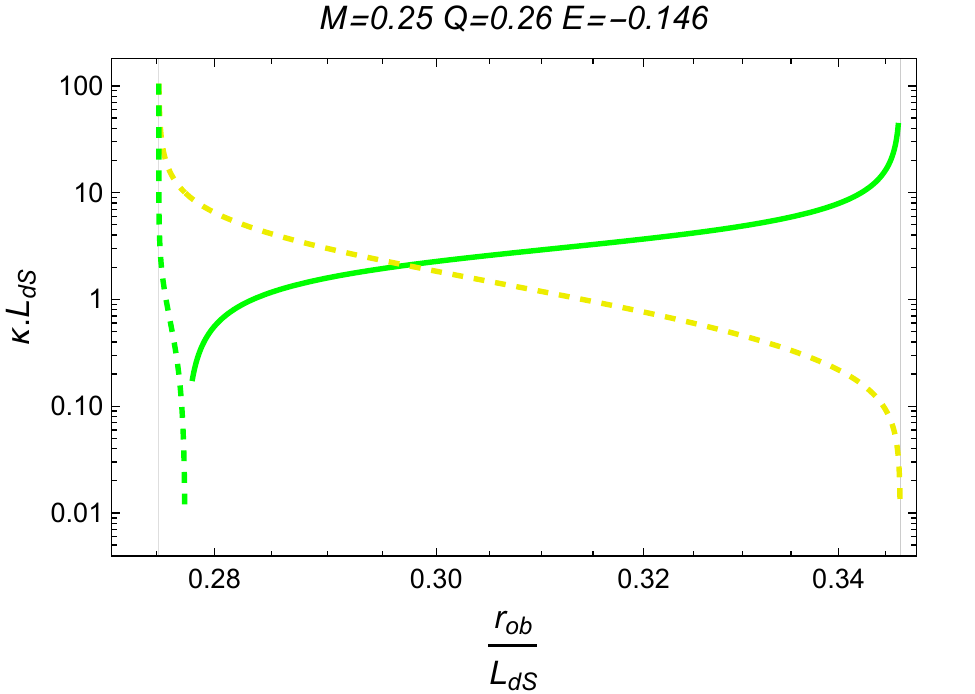}
    \end{minipage}%
    \vspace{1em}
    \begin{minipage}{0.49\textwidth}
    \centering
        \includegraphics[width=1\textwidth]{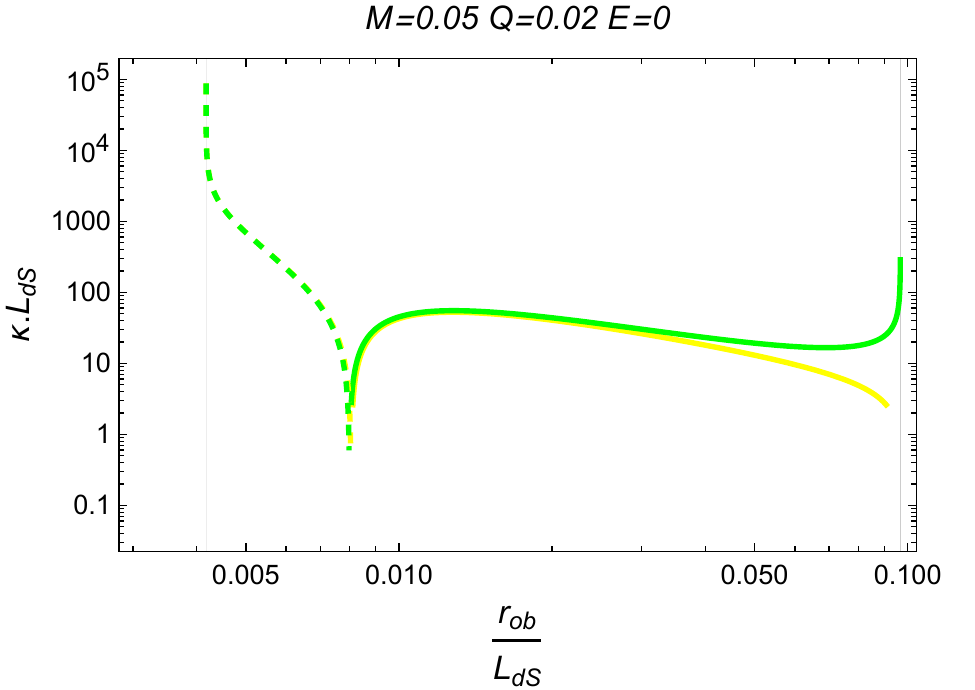}
    \end{minipage}%
    \caption{The first two graphs show the effective temperatures as seen by an observer with negative energy, plotted from $r_+$ to $r_-$ with gridlines at either horizon. The last graph depicts the effective temperature for an observer of zero energy. Green (yellow) represents the ingoing (outgoing) modes.}
    \label{fig:FIG 3}
\end{figure}

\begin{figure} [ht!]
    \centering
    \includegraphics[width=0.4\textwidth]{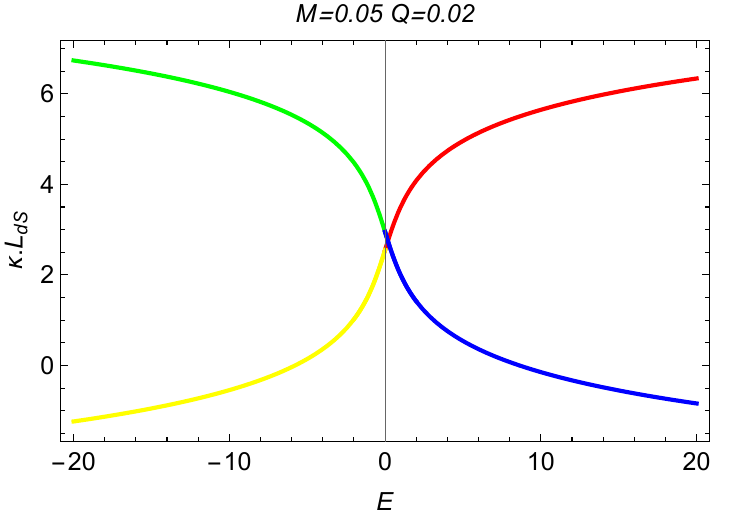}
    \caption{Log plot of the effective temperature as a function of the observer's specific energy at $r_{ob}=\frac{r_{-}+r_{+}}{2}$. The yellow (green) lines represent the effective temperatures of the outgoing (ingoing) modes for an outgoing observer (negative energy). The red (blue) lines correspond to the effective temperatures for an ingoing observer (positive energy).}
    \label{fig:NegEnergyFunc}
\end{figure}

\begin{figure} [ht!]
    \centering
    \includegraphics[width=0.5\textwidth]{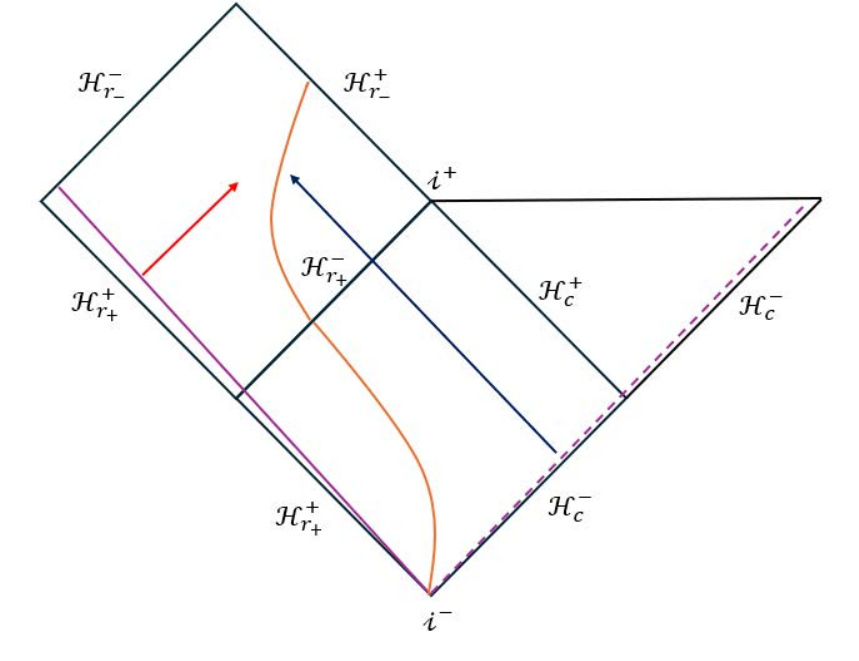}
    \caption{A Penrose diagram representing the trajectory of an observer with a negative specific energy beyond the event horizon. As in the diagram, the observer will eventually reach the right section of the inner horizon.}
    \label{fig:Penrose3}
\end{figure}

The effective temperatures perceived by an outgoing observer within the event horizon is shown in Fig.~\ref{fig:FIG 3}, and the Penrose diagram corresponding to the situation is given in Fig.~\ref{fig:Penrose3}. Such an observer will not see a very significant change in the Hawking modes, except for an interchange in the ingoing and outgoing effective temperatures. As a consequence, it will be the ingoing modes diverging to infinity at the inner horizon and the outgoing modes tending to a finite negative value. The apparent infinite blueshift in the ingoing modes may be attributed to an infinite Doppler shift as the horizon function goes to zero. But it is not only at the event horizon that this observer will see an infinite wall of energy\textemdash the horizon function similarly goes to zero at the event horizon, making it impossible for the observer to cross. This can be seen, once again, in the divergence of the ingoing modes at this region. The observer is therefore confined to the region between the event and the inner horizons, consistent with the nature of a trapping horizon.
An observer of zero energy is similarly unable to traverse beyond the event horizon. This situation is similar to the charged Nariai limit (in that the observer has zero energy), and indeed, there is thermal equilibrium for the most part, broken only as the observer approaches the event horizon. 
The cold regime also displays a divergence in the ingoing modes, with the outgoing modes starting to diverge as the energy becomes more negative. 
At any point inside the black hole, $\kappa_o^{-}\propto-E$ for $E<0$ and $k_i^+\propto E$ for $E>0$. Both $\kappa_i^-$ and $\kappa_o^+$ asymptotically tend to zero as $E\xrightarrow{}+\infty$ and $E\xrightarrow{}-\infty$ respectively, as in Fig.~\ref{fig:NegEnergyFunc}. Notably, the change in energy does not affect either the transition points or the sign in either of the modes. Neither the divergence at the inner horizon, nor the Hawking radiation itself can ever be forced to go to zero by a change in state of the observer.

In an RNdS black hole, there exists another trapping region outside the cosmological horizon. While this paper, in general, does not study this region in great detail, it is worth considering the implications of negative energy states in this region. The observer will be forced to move to a greater spatial radius (outfalling), but can be ingoing. The equilibrium in the modes as the observer tends to infinity remains unchanged, but the outgoing modes diverge as the observer approaches the cosmological horizon, while the ingoing modes tend to zero. Once again, this is consistent with the expectations that the observer should not be able to cross the cosmological horizon back into the observable universe. Unlike inside the event horizon, the effective temperatures in this region are independent of the specific energy in the infinite limit. However, the convergent distance for the ingoing and outgoing modes becomes farther with increasingly negative energy, and the divergence gets sharper toward the cosmological horizon. But once again this implies that it is not possible to Lorentz-boost an observer into a vacuum state that persists over all of spacetime for any parametric values of the black hole.

\section{Effective temperature perceived by radial free fallers in RNAdS spacetime(\texorpdfstring{$\Lambda<0$}{Lambda<0})} \label{sec:AdSTemp}

Having discussed, at great length, the effective temperatures perceived by an observer in de Sitter space, we now turn our attention to the anti-de Sitter  (AdS) space. AdS space has been long regarded as highly unrealistic, two reasons for which were cited by Hawking and Page \cite{HawkingPage_1983}: first, a negative cosmological constant corresponds to a negative energy density; second, closed timelike curves were seen to form on a manifold in the AdS space. While the second reason provided has been since invalidated \cite{Sokolowski_2016}, solving the apparent paradox by unwrapping the timelike circles and mapping them onto the real numbers ($-\infty$ to $+\infty$), the first reason still appears to stand. However, recent developments in quantum gravity proposed the holographic principle, a potential resolution for the information paradox. It involves a dimensional reduction to encode information from the bulk of an \textit{N} dimensional space onto the boundary of the space \cite{Hooft_2009}. Interest in the AdS space stems from the ability of the AdS/CFT correspondence to form a map from the bulk to the boundary, enabling the dimensional reduction \cite{Maldacena_1999,Rivelles_1999}.

Construction of the RNAdS spacetime is done by incorporating into the horizon function, a negative cosmological constant:
\begin{equation} \label{eq:horizonfnads}
    \Delta(r)=1-\frac{2M}{r}+\frac{Q^2}{r^2}+r^2,
\end{equation}
having  scaled by the anti-de Sitter length $L_{AdS}^{2}=\frac{-3}{\Lambda}$. It is a quartic polynomial in $r$, with only two real positive roots ($r_+$ and $r_-$), the larger of which represents the event horizon, the other representing the Cauchy horizon. The other two roots are negative and have no physical significance. The negativity of the cosmological constant eliminates the presence of a cosmological horizon. Furthermore, a simple examination of the horizon function reveals that it tends to infinity as $r$ increases (as does the first derivative), and is negative in between $r_+$ and $r_-$. 

The geodesic equations for this metric are not different from the RNdS geometry, so that the final equation for the frequency perceived either by the emitter or by the observer is still given by Eq.~\eqref{eq:omega}

\subsection{Vacuum state} \label{subsec:AdSVacuum}

A major distinction between the setup for the RNdS metric and the RNAdS metric comes from the fact that the vacuum state cannot be defined as easily in a contracting universe as it is in an expanding one. The Unruh vacuum, as described earlier, requires that the outgoing modes be positive with respect to the past horizon, and the ingoing modes be positive frequency with respect to $\partial/\partial t$ at past null infinity which would generally be the outer boundary of the spacetime. But this boundary is infinity, which gets compressed as the space contracts. Since $\Delta\xrightarrow{}\infty$ as $r\xrightarrow{}\infty$, a clock on any emitter placed at this boundary will have to tick infinitely faster than the global time coordinate. So while the Unruh vacuum can be defined with respect to the past horizon, it is ill defined at the AdS boundary. The eternal AdS geometry therefore cannot be modeled in entirety by the Unruh vacuum. 

To construct the vacuum state, we will consider a family of emitters in free fall at the event horizon. The outgoing modes emitted are of the form (once again incorporating the geometric optics approximation)
\begin{equation}
    \phi^{out}=\frac{1}{\sqrt{4\pi \omega}}e^{-\varkappa_+ U_+},
\end{equation}
where the usual null coordinates are defined as earlier in Eq.~\eqref{eq:EFcoordinates}

A very similar proof as in Section~\ref{subsec: UnruhState} can be used to show that the proper time of an emitter in free fall at the event horizon is proportional to the outgoing Kruskal coordinate at the event horizon. The outgoing modes will travel out to the boundary of the AdS spacetime at infinity, from where they will be reflected back as ingoing modes. The AdS spacetime may be viewed as a finite box, which will slowly get filled with radiation \cite{Hemming_2001}. At some point in process, a thermal equilibrium is reached, where the ingoing and outgoing modes have the same effective temperature. The configuration is modeled best by the Hawking-Hartle state, where the modes are positive with respect to the affine parameters on their respective boundaries of generation. \cite{Balasubramanian_1999} offers a more detailed insight into the mode solutions in AdS spacetime.

\subsection{Effective temperature}

\begin{figure} [ht!]
    \centering
    \includegraphics[width=0.5\textwidth]{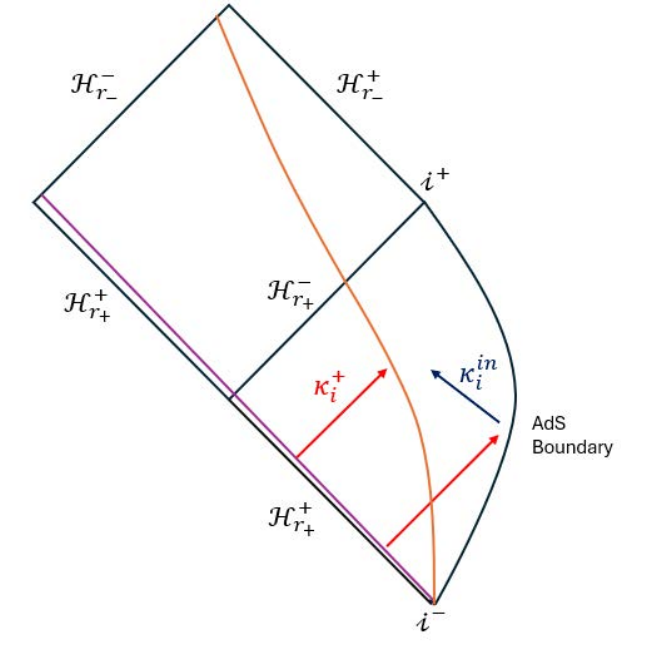}
    \caption{Both the observer (orange) and the emitter (purple) are ingoing, with the observer having been released from an arbitrary point well outside the event horizon. Outgoing modes are always produced from the ingoing family of emitters, but may reflect off the AdS boundary to yield the ingoing modes.}
    \label{fig:Penrose2}
\end{figure}

\begin{figure*}
    \centering
    
    \begin{minipage}{0.49\textwidth}
    \centering
        \includegraphics[width=1\textwidth]{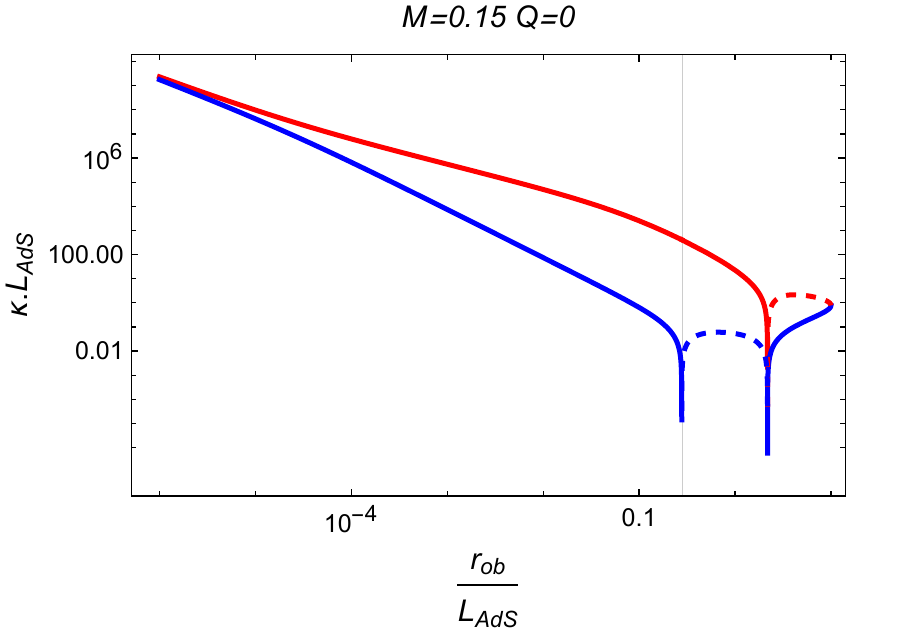}
    \end{minipage}
    \hfill
    \begin{minipage}{0.49\textwidth}
    \centering
        \includegraphics[width=1\textwidth]{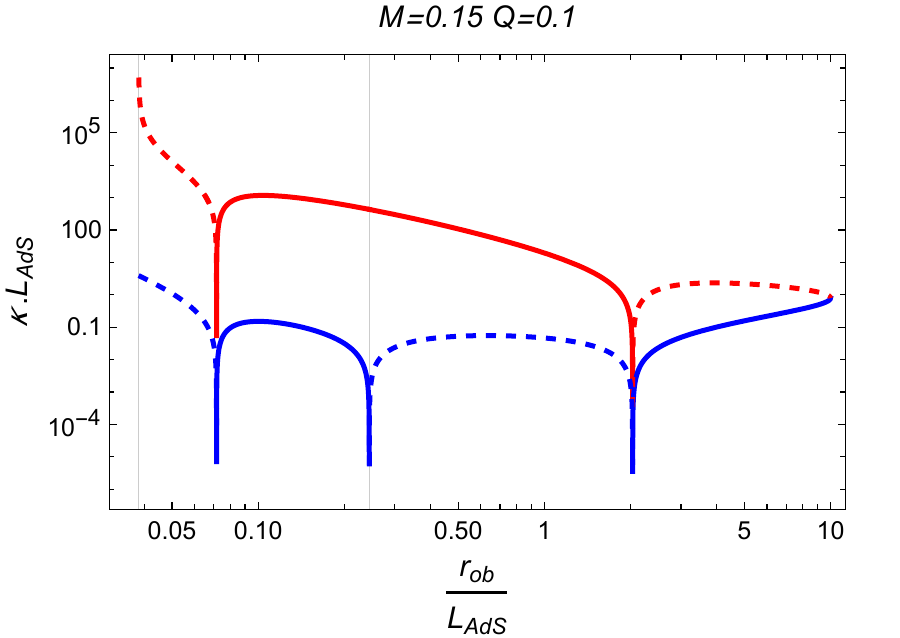}
    \end{minipage}%
    \vspace{1em}
    \begin{minipage}{0.49\textwidth}
    \centering
        \includegraphics[width=1\textwidth]{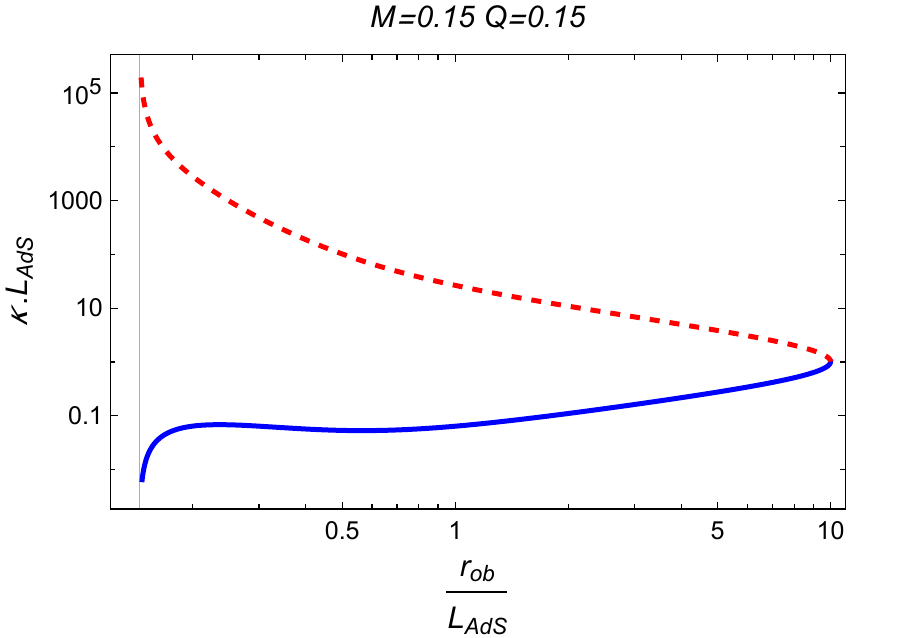}
    \end{minipage}
    \hfill
    \caption{Effective temperatures for a free falling observer in AdS space. Red curves represent $\kappa^+$ and blue curves represent $\kappa^{-}$. Gridlines are at the Cauchy and event horizons. At \textit{Q}=0 the inner horizon does not exist. At M=0.15,Q=0.15 the two horizons coincide.}
    \label{fig:AdS}
\end{figure*}

In Section~\ref{subsec:efftempfunc} we had derived four effective temperature functions corresponding to the outgoing/ingoing modes and outgoing/ingoing observers. In AdS spacetime, an observer released from rest at any point will always be infalling (unless subsequently provided with sufficient energy),
so that there are only two forms of Eq.~\eqref{eq:kappanew}:
\begin{subequations}
\begin{align}
    &\kappa_i^{+}(r_{ob}) = -\frac{E+\sqrt{E^2-\Delta(r_{ob})}}{\Delta(r_{ob}) L_{AdS}}\left(\varkappa(r_{ob}) -\varkappa_{+}\right),\\
    &\kappa_i^{in}(r_{ob}) = \frac{E-\sqrt{E^2-\Delta(r_{ob})}}{\Delta(r_{ob}) L_{AdS}}\left(\varkappa(r_{ob}) -\varkappa_{+}\right).    
\end{align}
\end{subequations}

A change in notation from superscript \textit{c} to ``in'' has been introduced to reinforce the fact that the ingoing modes are no longer generated from the cosmological horizon. The only difference between these equations and the ones for a dS spacetime is that the equation for the ingoing modes depends upon $\varkappa_+$ instead of $\varkappa_c$. The redshift experienced by the modes on the way out from the position of the observer to infinity is canceled by a corresponding blueshift as it reflects and travels back, as shown in Fig.~\ref{fig:Penrose2}. The net red/blueshift observed is only that which occurs between the event horizon and the position of the observer. It is akin to the ingoing modes having been generated from the event horizon itself. The equations implicitly assume that the outgoing modes are radial in the spherical geometry, and undergo perfect reflection at the AdS boundary with a zero phase shift. 

Having determined the nature of  the modes, we must next address the state of the observer/emitter. Generally, we would choose to release the observer from rest at the maxima of the horizon function. That would also conclusively determine the specific energy of the observer, which remains constant throughout the trajectory. Given that this horizon function tends to positive infinity, such a choice cannot be made. The only way for the observer to start out from infinity and not have an infinite specific energy would be to have an infinite radial velocity, which would only increase further along the geodesic. The most sensible option is then to release the observer from rest at any point outside the black hole. The metric provides no other points of geometric interest outside the event horizon. Recalling that we are working in units of anti-de Sitter length, we can choose a point far enough that it covers all regions of interest while also giving a value of the specific energy that is not exceedingly large. The value must furthermore, be large enough to include the inflection point in the effective temperature, which occurs when the slope of the horizon function at $r_{ob}$ exceeds that at the event horizon. A value of $r_0=10$ (AdS length units) has been chosen as a standard rest point, unless stated otherwise. 

The simplest case of an RNAdS black hole, is once again, when the charge goes to zero, which falls under Schwarzschild-anti-de Sitter (SAdS) limit, the graph of which is on the upper left of Fig.~\ref{fig:AdS}. Having analyzed in Section~\ref{sec:efftemperature}, the effects of the mass and positive cosmological constant, the SAdS limit provides the perfect context in which to examine the effects of the cosmological contraction, in isolation from the effects of mass and charge. Knowing that the mass causes a redshift, the negative effective temperatures in the outgoing modes may be attributed entirely to the effects of the negative cosmological constant. The same assumption cannot be made for the ingoing modes, given that their source of origin is not the same as it was in the SdS case. The effects of this will be analyzed separately. As the observer starts out from rest and falls inward, traveling in predominantly AdS space, the outgoing modes appear blueshifted, with the rate of blueshift decreasing as the effects of mass takeover. It is not unexpected that if $\Lambda>0$ results in a redshift, that $\Lambda<0$ causes a blueshift. The outgoing modes are decreasingly redshifted away from the event horizon, and become increasingly blueshifted as they travel out to infinity. Having reflected off the boundary, they then become ingoing modes. At this point, the frequency of the modes are traversing a slope of the horizon function greater than that at the event horizon, resulting in a gravitational redshift. In addition, the observer is also ingoing, resulting in a Doppler redshift as well. Effectively, it is the source of the ingoing modes that results in the perceived redshift in AdS space. As they near the event horizon, they start to experience a gravitational blueshift. Like in the previous cases, the point of transition from positive to negative effective temperatures occurs where $\varkappa(r_{ob})=\varkappa_+$. Because both the ingoing and outgoing modes source from $r_+$, they both undergo transitions at exactly the same points, except for the transition of the ingoing modes at the event horizon (a consequence of a zero Doppler shift). The mass predominates beyond the event horizon, and the graph, in this regime, bears close resemblance to those of the SdS and Schwarzschild cases \cite{McMaken_2023}, where the effective temperatures are positive and eventually converge, with the sharpness of the convergence being proportional to \textit{M}. There are also no parameters at which the SAdS black hole exists in thermal equilibrium, and except for at the rest point, the observer never perceives an equal effective temperature (magnitude) in the ingoing and outgoing modes. This is once again a result of the family of emitters for the ingoing modes (the Doppler shift factors are never equal, but the gravitational redshift factors are always equal).

The upper right graph of Fig.~\ref{fig:AdS} represents a fairly general case of the RNAdS parameter space. The properties of the inner horizon are the same as in the RNdS metric, i.e.\ it will have a negative surface gravity, thereby repelling the infalling modes. The divergence of the outgoing modes at the horizon, and the blueshift in the ingoing modes, therefore remains the same. While the general behavior of the modes up to the blueshift at the inner horizon does not change, the relative distances of the transition points change depending on the cosmological constant and the $Q/M$ ratio. 
As the cosmological contraction increases ($\Lambda$ decreases by becoming more negative, and the scaled mass parameter $M$ increases), the outer transition point comes closer to the event horizon as would be expected if the AdS nature of spacetime takes precedence over mass closer to the black hole. As the inner and event horizons approach each other, it is possible for the ingoing modes to continue as a negative value across the event horizon, approaching the degenerate limit of \textit{M}= \textit{Q} as in lower graph of the figure. This critical ratio for this continuity decreases as the cosmological constant gets more negative. The extent of the intermediate redshift (between $r_-$ and $r_+$) is seen to decrease as the \textit{M}/\textit{Q} ratio decreases, once again consistent with the fact that a larger charge parameter should contribute to a greater blueshift.

\section{Validity of the adiabatic approximation} \label{sec:Adiabaticity}

\begin{figure*}
    \centering
    \begin{minipage}{0.49\textwidth}
    \centering
        \includegraphics[width=1\textwidth]{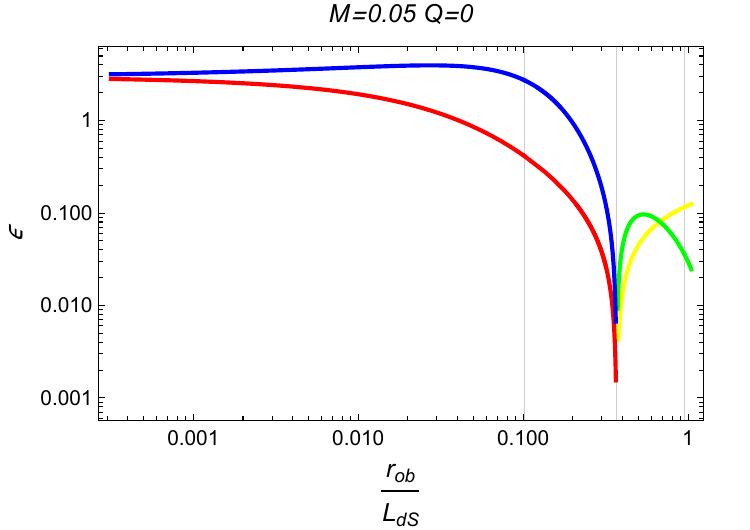}
    \end{minipage}
    \hfill
    \begin{minipage}{0.49\textwidth}
    \centering
        \includegraphics[width=1\textwidth]{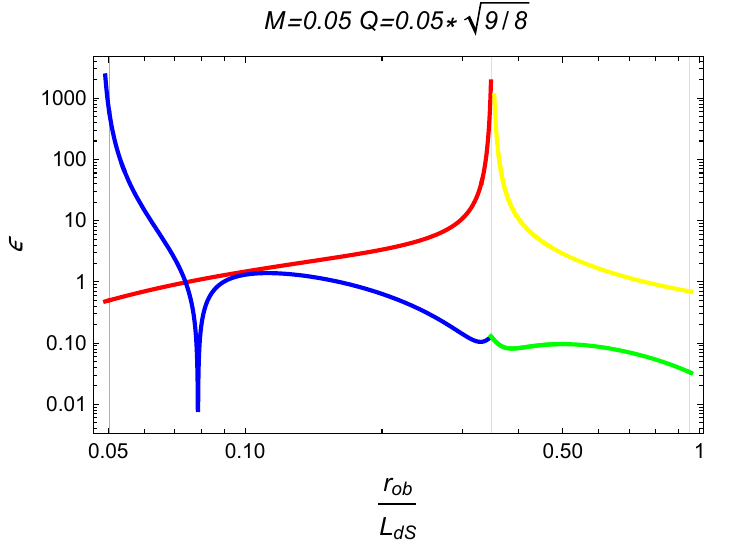}
    \end{minipage}%
    \caption{Adiabatic control function against the radius of the observer in RNdS space.}
    \label{fig:AdiabaticdS}
\end{figure*}

\begin{figure} [ht!]

    \centering
        \includegraphics[width=1\columnwidth]{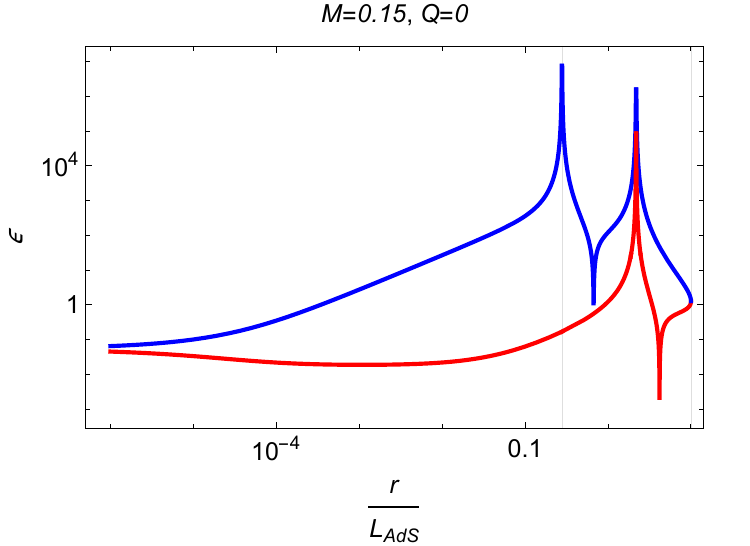}

    \caption{Adiabatic control function against the radius of the observer in SAdS space.}\label{fig:AdiabaticAdS}
\end{figure}

The validity of the effective temperature formalism relies upon the fulfillment of the adiabatic condition, which requires that the frequency of an emitted photon (and therein $\kappa$) remains approximately constant over one oscillation of the electromagnetic field. Satisfaction of the adiabatic condition is sufficient to guarantee the existence of a Planckian distribution of thermal Hawking radiation.
The adiabatic control function is defined as \cite{Barcel__2011}
\begin{equation}
\begin{split}
    \epsilon&=\left|\frac{\dot{\kappa}(u)}{\kappa(u)^2}\right|\\
    &=\frac{1}{\kappa(u)^2}\left|\frac{d\kappa}{d\tau_{ob}}\right|\\
    &=\frac{1}{\kappa(u)^2}\left|\frac{\partial \kappa}{\partial r_{ob}}\dot{r}_{ob}+\frac{\partial \kappa}{\partial r_{rem}}\dot{r}_{em}\frac{\omega_{ob}}{\omega_{em}}\right|,
   \end{split}
\end{equation}
and must satisfy $$\epsilon\ll 1.$$
The overdot represents the derivative with respect to the proper time of the observer/emitter.
Neither the analytic forms of the control function, nor its limit under any special circumstances are straightforward enough to be directly enlightening. However, the general behavior of $\epsilon$ may be inferred from the graphs in Figs.~\ref{fig:AdiabaticdS} and \ref{fig:AdiabaticAdS}. The superscript convention for the control function $\epsilon$ will follow that introduced for the corresponding effective temperature functions in both the RNdS and RNAdS models.

Adiabaticity is maintained to the greatest extent for an outfaller beyond the rest point. $\epsilon$ is continuous across the cosmological horizon and into the limit of a predominantly de Sitter space. Generally, the control function goes to a minimum at the rest point of the observer ($\dot{r}_{ob}=0$). The exception is in the extremal case (right panel in Figure~\ref{fig:AdiabaticdS}),  where the control function diverges for outgoing modes, due to the transition in the effective temperature of the outgoing modes occurring exactly at the degenerate horizons ($r_0=r_+=r_c$). $\epsilon$ diverges as the effective temperatures go to zero; the formalism is therefore least valid near the transition points.
For general RNdS models, both $\epsilon^+$ and $\epsilon^c$ are close to the zeroth order of magnitude inside the event horizon and rise sharply at the transition points before coming back down the zeroth order close the Cauchy horizon. But the zeroth order of magnitude is still greater than one, indicating that adiabaticity is not maintained. There are therefore, only three regions in which the approximation holds: at and beyond the rest point, slightly before the transition points where $\frac{d\kappa}{dr_{ob}}\approx0$ and close to the inner horizon. In the SdS limit, however, there being no inner horizon, $\lim_{r\to 0} (\epsilon^{+},\epsilon^{c})=3$. The control function is only less than one in the proximity of the rest point.

In anti-de Sitter space, the ingoing modes are, in general, adiabatic in only two regimes: first is near the event horizon, at the minima of the effective temperature function, second is at the local maxima before the inner horizon. For an SAdS model, both $\epsilon^+$ and $\epsilon^{in}$ converge as they approach the singularity. The asymptotic limit is proportional to the mass parameter, and is less than 0.5 for $M<2$. As the mass parameter increases, the cosmological constant becomes less negative (i.e.\ increases), and the contraction reduces. As the contraction reduces, adiabaticity is maintained to a greater extent inside the event horizon. 
It also generally appears to be true that the outgoing modes are more adiabatic than the ingoing modes. This is consistent with the graphs from Fig.~\ref{fig:AdS}, according to which the magnitude of the outgoing modes is greater than the ingoing modes. 

For the most part of the observer's trajectory, the adiabatic condition does not hold. One may question the implications for the interpretation of $\kappa$. The form of $\kappa$, as in Eq.~\eqref{eq:kappanew} and used throughout the paper, is derived directly from the definition given by Eq.~\eqref{eq:kappa}. It does not rely upon the adiabatic approximation. The observer will still see a nonzero particle production, and therein the Hawking radiation. But the radiation will not have a thermal blackbody spectrum. The graphs in Figs.~\ref{fig:AdiabaticdS} and \ref{fig:AdiabaticAdS} show the regime in which the radiation is thermal.  

\section{Conclusions}

There are four main extremities in the RNdS parameter space: Schwarzschild-de Sitter, Nariai, the lukewarm family and the extremum of the cold regime. An observer of positive energy was released from the maximum of the horizon function, either infalling toward $r_+$ or outfalling toward $r_c$. The outfaller always sees positive effective temperatures, eventually finding an asymptotic state of thermal equilibrium in pure de Sitter space. The infaller generally perceives positive effective temperatures until they reach a point in between $r_+$ and $r_-$ where the Cauchy  horizon induces a blueshift, causing the observer to perceive a negative effective temperature. An exception is at the extremal cold limit, where the observer never sees a negative effective temperature in the ingoing modes. For values of the charge that are large enough, it is possible for the observer to see a blueshift in the outoing modes before they reach the event horizon. The outgoing modes are always negative and generally diverge to infinity at the inner horizon.

To reinforce the generality of the divergence, a density plot of $\log(\kappa^{+}_i\Delta)$ evaluated at the inner horizon was plotted over the entire permissible parameter range of the RNdS metric. A finite value of $\kappa^{+}_i (r_-)$ causes the product $\kappa^{+}_i\Delta$ to go to zero. The product was finite and nonzero over the interior of the sharkfin indicating that the effective temperatures were negatively divergent. However, in accordance with semiclassical predictions, the effective temperature at the inner horizon was seen to be finite for cold and ultracold black holes. Differing from the predictions of \cite{Cardoso_2018}, the effective temperature still diverges to infinity in the lukewarm regimes. The effective temperature formalism is seen to be one step more accurate in locating regions of possible violation than an examination of the quasinormal modes at the inner horizon. Yet, it is still a semiclassical formulation that does not have the robustness of the quantum stress-energy tensor approach, which further narrows down the violation to the ultracold regime.  

The effective temperatures observed by free fallers of negative specific energies were similarly graphed. The ingoing and outgoing modes exchange their properties, so that the ingoing modes diverge at the inner horizon, while the outgoing modes are finite. The divergence of the ingoing modes at the event horizon reinforces that observers of negative energy may exist only inside the trapping region. Similar observations were made for negative energy observers outside the cosmological horizon, with the outgoing modes diverging at $r_c$. But it remains true that an observer approaching the inner horizon will encounter a wall of infinite energy, making it impossible to cross. It is also not possible to Lorentz-boost an observer into a frame where either the Hawking radiation or the divergence is eliminated. 

In contrast to a cosmological expansion is the case of a cosmological contraction, with a black hole set in an anti-de Sitter space. Unlike for dS spacetime, AdS features no cosmological horizon, with its outer boundary being infinity. Outgoing modes are sourced from the past horizon, reflect off the boundary at infinity and travel back inward as ingoing modes. An observer, starting from an arbitrarily chosen rest point will see the outgoing modes get blueshifted under the effect of the contraction, with the mass of the black hole taking predominance, slowly causing the modes to redshift. The outgoing modes remain redshifted until they once again undergo a blueshift effect under the influence of the Cauchy horizon. Yet again, there is an infinite divergence at the inner horizon. 

Finally, the validity of the adiabatic approximation was tested by plotting the adiabatic control function against the position of the observer. It is seen, that in general, adiabaticity does not hold, except around a select set of regions. An observer in regimes where adiabaticity is not fulfilled  will still see radiation\textemdash but it will not be a thermal black body spectrum. While the function $\kappa(r)$, is not affected, the physical conceptualization that accompanies it is no longer that of a ``temperature'' in conventional sense. But the observer will still see  nonzero particle production. A more rigorous calculation of the particle spectrum observed, in regimes where the spectrum is nonthermal, may be accomplished by calculating the scattering coefficients of the modes as seen by an observer in the vacuum state of the emitter \cite{McMaken_2024}. This calculation, while not covered in this paper, is a worthwhile attempt reserved for future work. 

\bibliography{bibliography.bib}

\end{document}